\documentclass[prd,twocolumn,showpacs]{revtex4}

\usepackage{epsfig}
\usepackage{graphics}

\begin{document}

\title{Precision of Hubble constant derived using black hole binary absolute distances and statistical redshift information}
\author{Chelsea L. MacLeod}
\author{Craig J. Hogan}
\affiliation{University of Washington, Department of Astronomy, Box 351580\\
Seattle, WA~~98195-1580}
\date{\today}
\begin{abstract}
Measured gravitational waveforms from black hole binary inspiral
events  directly determine absolute luminosity distances. To use these
data for cosmology, it is necessary to independently obtain
redshifts for the events, which may be difficult for those without
electromagnetic counterparts. Here it is demonstrated
that certainly in principle, and possibly in practice,
clustering of galaxies allows extraction of the redshift information
from a sample statistically for the purpose of estimating mean
cosmological parameters, without identification of host galaxies for
individual events.  We  extract mock galaxy samples from the 6th Data
Release of the Sloan
Digital Sky Survey  resembling those that would be  associated with
inspiral events of stellar mass black holes falling into massive black
holes at redshift $z\approx 0.1$ to $0.5$. A simple statistical
procedure is described to estimate a likelihood function for the
Hubble constant $H_0$: each galaxy in a LISA error volume contributes
linearly to the log likelihood for the source redshift,  and the log
likelihood for  each source contributes linearly to that of $H_0$.
This procedure is shown to provide an accurate and unbiased estimator
of $H_0$. It is estimated
that a precision better than one percent in $H_0$ may be possible if
the rate of such events is sufficiently high, on the order of 20 to $z=0.5$.

\end{abstract}
\maketitle
\section{Precision cosmology from black hole binaries}
A low frequency gravitational wave detector,  such as LISA, is capable
of measuring very high  signal to noise  waveforms from inspirals and
mergers of cosmologically  distant black hole binaries.  From the
measured waveform alone it is possible to estimate the parameters of a
binary with high precision, including its direction on the sky and its
absolute luminosity distance \cite{1986Natur.323..310S}. Aside from numerical factors, the absolute radius of the final hole is fixed by the square of the orbital period divided by the  orbit decay or chirp time; the distance is this absolute length divided by the measured wave amplitude. 
The gravitational  calibration of distance  is not accompanied by the usual systematics associated with astronomical modeling; indeed it does not even require Standard Model physics.
 Since a single black hole binary merger can provide distances with
 absolute precision much better than one percent, this capability may
 offer a potentially transformative tool for precision measurement of
 cosmological parameters
 \cite{2006PhRvD..74f3006D,2005ApJ...629...15H} (for reviews on
 gravitational waves see Ref.s \cite{flan,2007arXiv0709.0608H,2007CQGra..24..113A,LISA}).  Precision measurement of an absolute distance scale, as embodied in the Hubble constant, is important for breaking degeneracies in estimates of cosmological parameters (such as those characterizing cosmological curvature and Dark Energy) using other datasets, such as cosmic background radiation anisotropy \cite{Hu,Knox}.

On the other hand there are obstacles in practice to applying the
technique.  One problem is distance errors added by  gravitational
lensing along the line of sight
\cite{2003CQGra..20S..65H}.
The best raw distances are given by massive black hole (MBH) binary
inspiral events (where both holes are larger than $10^4M_\odot$ say),
many of which will be measured with very high signal-to-noise ratio \cite{2007arXiv0707.4434T,2005ApJ...629...15H,2005PhRvD..71h4025B,2004PhRvD..70d2001V,2002MNRAS.331..805H,1998PhRvD..57.7089C}.
They are predicted to be fairly frequent (one or two events per week
on average) and should be observable with LISA out to redshifts
greater than 10 \cite{2004ApJ...611..623S,2006AIPC..873...61V}.
Unfortunately most of the observable events are predicted to occur at
redshifts greater than 2 where errors due to lensing are substantial. 
Thus one is led to consider another class of events, stellar mass
black holes inspiralling into massive black holes (the so-called
extreme mass ratio inspiral or EMRI events
\cite{2006PhRvD..74b3001D,2004CQGra..21S1595G,2004PhRvD..69h2005B,2006AIPC..873..241H,2006ApJ...645L.133H}), which occur frequently in
galaxies at $z<1$ where lensing errors are subdominant. Although the
intrinsic precision of these distances is lower due to lower
signal-to-noise ratio, they still may provide a unique and precise cosmological probe.

The other major uncertainty is associated with measuring the redshifts
of the events.  The gravitational waveform provides a measurement of
luminosity distance, and a cosmological probe   such as a mean
redshift-distance relation requires an independent measurement of
redshift. In the case of MBH events, it may be possible to identify
the host galaxy by seeking electromagnetic counterparts \cite{2006MNRAS.372..869D,2005ApJ...622L..93M,phinney,2002ApJ...567L...9A,2006ApJ...637...27K,2007PhRvD..76b2003K,2007arXiv0712.1144K} to the merger,
such as optical or x-ray variability in accretion disks from the rapid
evolution in the gravitational potential of the binary black hole, or
even from the sudden removal of several percent of $Mc^2$ in
gravitational radiation at the moment of merger. But in the case of
EMRI events, no compelling model requiring an electromagnetic counterpart
has been offered: it may be possible to merge a small black hole with
a large one with practically no signature aside from the gravitational radiation itself.

The point of this paper is to sketch and demonstrate a technique for
measuring cosmological parameters such as the Hubble constant $H_0$ by
obtaining only statistical information about the redshifts of the EMRI
hosts, even without identifying the host galaxy of any individual
event.  For each event, the estimate of direction $\vec\theta$ and
distance $D_L$, together with a cosmological model relating $D_L$ and
redshift $z$ roughly estimated from other techniques, provide a
three-dimensional ``error box'' in $\vec\theta, z$ space. As first
noted in \cite{1986Natur.323..310S}, a galaxy redshift survey in this
error box then provides a statistical estimate of the host redshift,
since galaxies are highly clustered with each other. Here we demonstrate using a real survey 
a technique for estimating the likelihood distribution of the host
galaxy redshift, which is more accurate (statistically) than the prior information on $H_0$ that went into constructing the error box.   

In order to demonstrate the utility of this technique in practice, we
construct mock LISA error boxes in the Sloan Digital Sky Survey (SDSS) 
volume \cite{DR6,2002AJ....123..485S,2002AJ....124.1810S,2006AJ....131.2332G,camera,2000AJ....120.1579Y,filters} and generate
mock redshift surveys from samples of SDSS galaxies that have about
the same statistical clustering properties as a host galaxy population
of LISA EMRI sources.  One reason to choose a real galaxy survey
rather than a simulation for these realizations is that the higher
order correlations between galaxies in the cosmic web, which are
important for determining the likelihood distribution of redshifts,
are known to be correct, as long as the SDSS catalog selection
approximates an unbiased sample of a typical EMRI host galaxy
population.  We find that with enough events, precision better than one percent is possible in measuring the mean redshift-distance relation.
This translates into comparable precision in $H_0$, by a technique that shares very few systematic errors or biases with other means of measuring $H_0$.   It also enables other new cosmological tests, for example,  direct measurement of cosmic acceleration within the redshift range ($z<0.5$) where the universe is dominated by Dark Energy. 
 
\section{Use of statistical redshifts for cosmological parameter estimation from gravitational waveforms}
A fit to  a gravitational wave signal from a black hole inspiral event $j$
leads to a likelihood distribution   $\ln{\cal L}_j(\vec\theta, D_L)$ for its angular location  $\vec\theta$ and luminosity distance $D_L$. 
We wish to combine this with information from the directions and redshifts of galaxies to measure mean cosmological parameters such as the Hubble constant $H_0$.

For the simple realizations shown in Section III., the selection function is highly idealized: all galaxies are assumed to be equally likely hosts,  within a certain error box: the angular size of the box is defined by LISA errors in angle,  and the depth of the box is determined primarily by prior errors on $H_0$ from other sources. For this  discussion, intended only to estimate typical errors,  we also assume linear Hubble flow, in particular negligible cosmic acceleration (though this is taken into account where needed in the scaling exercise below; the linear discussion  is valid as long as acceleration is negligible within each redshift error box).
With this simple selection and expansion model, 
the log-likelihood distribution for the Hubble constant for each event
$j$ is (up to a constant numerical factor)
\begin{equation}
\ln{\cal L}_j(H_0)=N_j^{-1}\sum_i \ln{\cal L}_j(D_j=cz_i/H_0),
\end{equation}
summed over the galaxies $i$ in the box, and for a whole sample,
\begin{equation}
\ln{\cal L}(H_0)
=\sum_i \sum_j N_j^{-1} \ln{\cal L}_j(D_j=cz_i/H_0),
\end{equation}
where a  normalization factor $N_j$, the number of galaxies actually
measured in each box, is included so that each event is weighted
equally independent of the number of galaxies measured.  Thus each
galaxy gets to vote appropriately on its source redshift, while each
EMRI source contributes on an equal footing to estimating a value for $H_0$.

If the redshift distribution were very smooth on the scale of the LISA
error boxes, this technique would  not return useful information on
redshift: the likelihood distribution of $H_0$ would be  the same as that going into construction of the error boxes.  However, actual galaxies are highly clustered so if the angular errors are not too large, there is recoverable statistical information on $z$ as demonstrated quantitatively below.

There is a possibility of some bias in the estimate of $H_0$, for example if the rate of EMRIs is rapidly evolving with redshift, which means the weighting within each box should not be uniform in $z$. Since the mean evolution will actually be measured, such biases can be measured and accounted for statistically; however, we do not  model them here.

\section{Mock LISA/EMRI host-galaxy samples from SDSS}
If galaxies are sufficiently clustered,  then within the region of
$\vec\theta_i,D_i$ allowed for each event, a redshift catalog can give an estimate of the host redshift without knowing which galaxy is the host.
The question is whether for realistic galaxy clustering and LISA error
boxes, there is enough redshift information to be useful. We answer this question by generating mock redshift samples from regions of the SDSS volume scaled assuming standard cosmology to have about the same clustering properties as EMRI hosts.

We use estimates of LISA errors in distance and sky position from Cutler  (private communication). For  
$10+ 10^6M_\odot$ EMRIs at a redshift $z$, and a fully functioning LISA with  two effective  synthetic interferometers,
we use an error box with a solid angle $\Delta \theta^{2} \approx 4z^{2}$ square  degrees,
and  assume distance error $\Delta \ln(D_L) \approx 0.05 z$.  For a
LISA with only one synthetic interferometer, we adopt $\Delta
\theta^{2} \approx 16 z^{2}$ square  degrees, and   $\Delta \ln(D_L) \approx 0.07 z$. It should be noted that these
estimates only define a 63\% confidence interval; therefore, in a
proper simulation one should increase the size of the error box by
some amount to improve the confidence limits. However, in our exploratory
analysis we neglect this correction.

\begin{figure*}[htbp]
   \centerline{
     \includegraphics[width=2.2in]{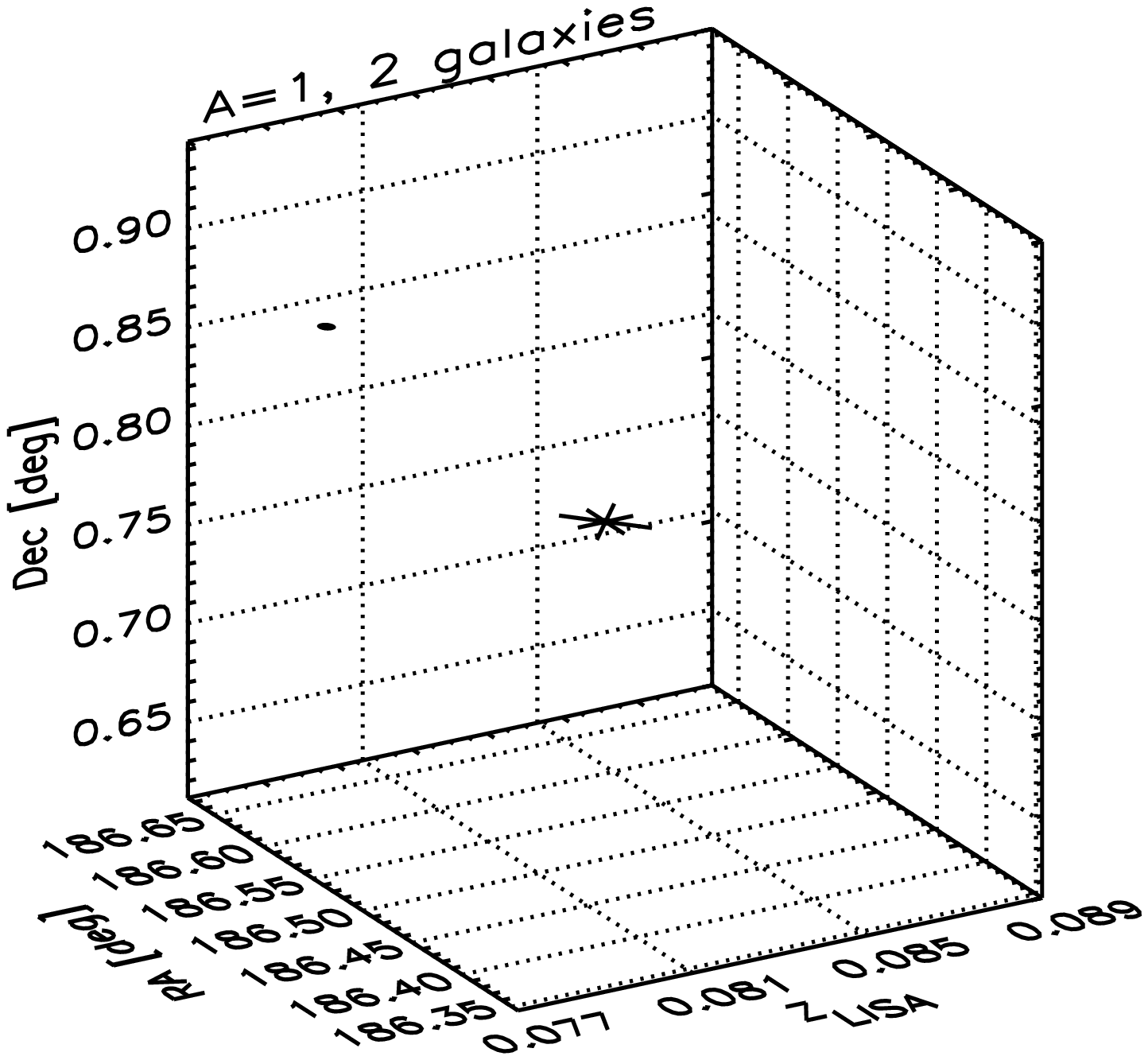}
     \includegraphics[width=2.2in]{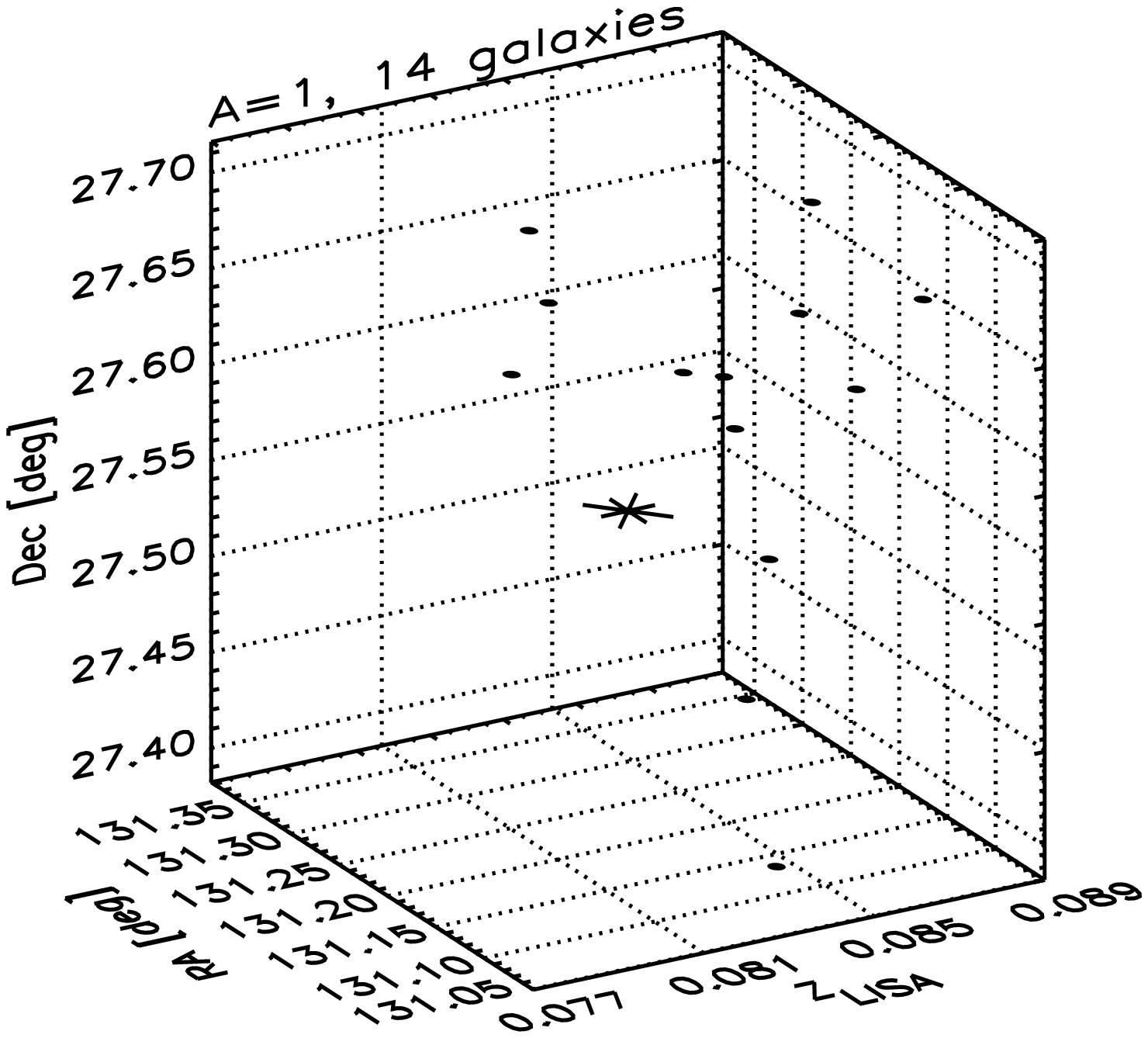}
   }\centerline{
     \includegraphics[width=2.2in]{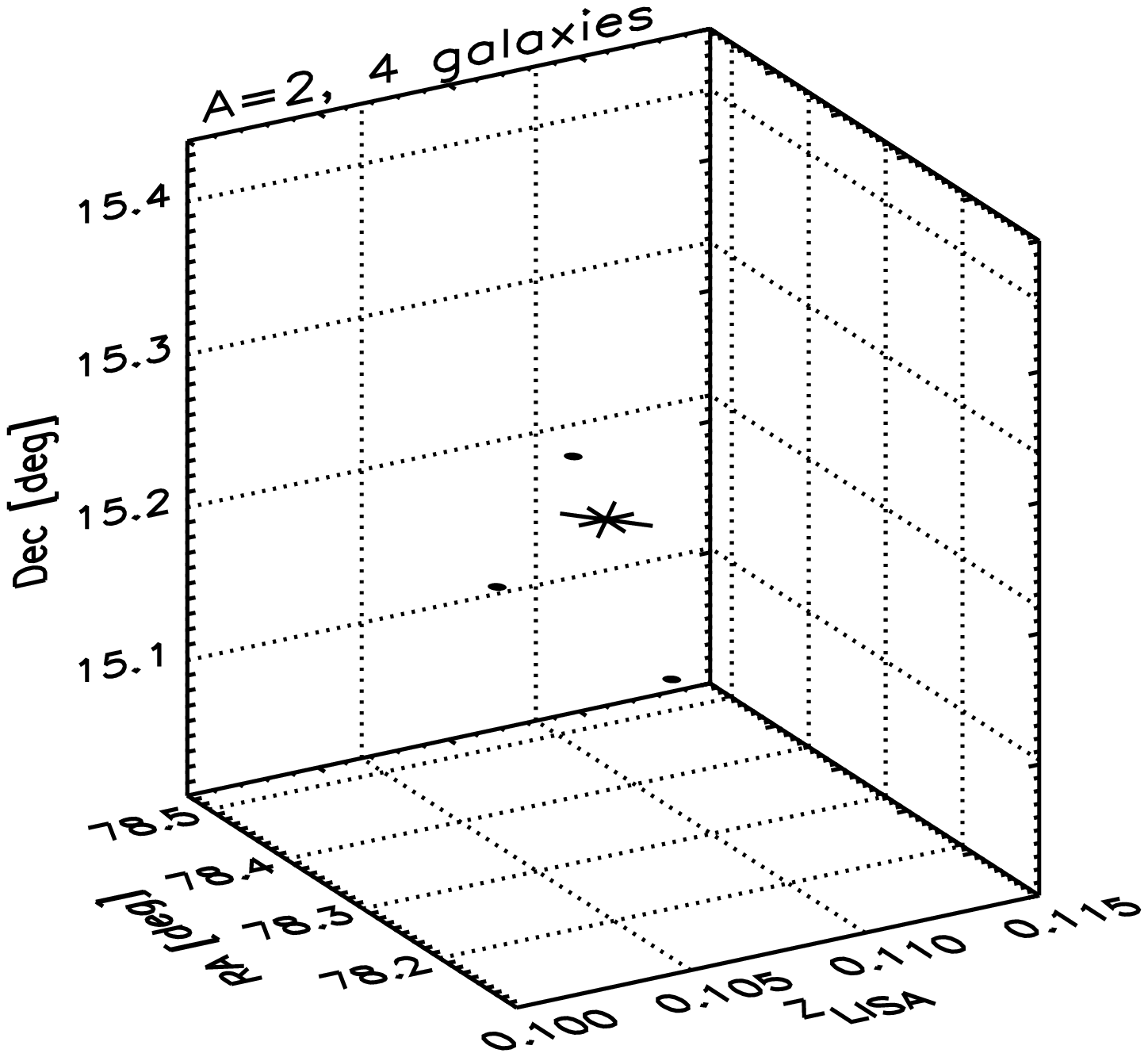}
     \includegraphics[width=2.2in]{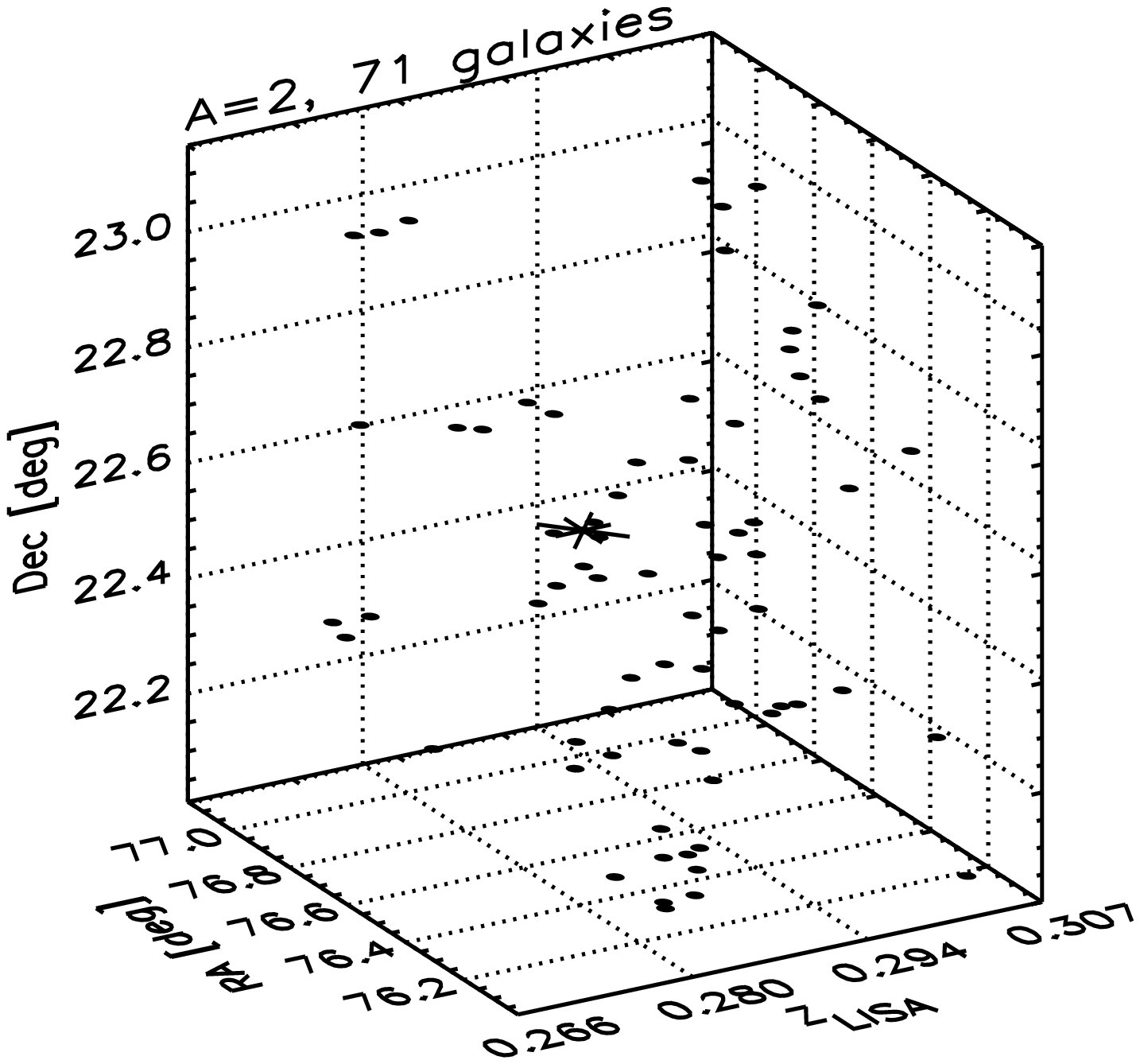}
   }\centerline{
     \includegraphics[width=2.2in]{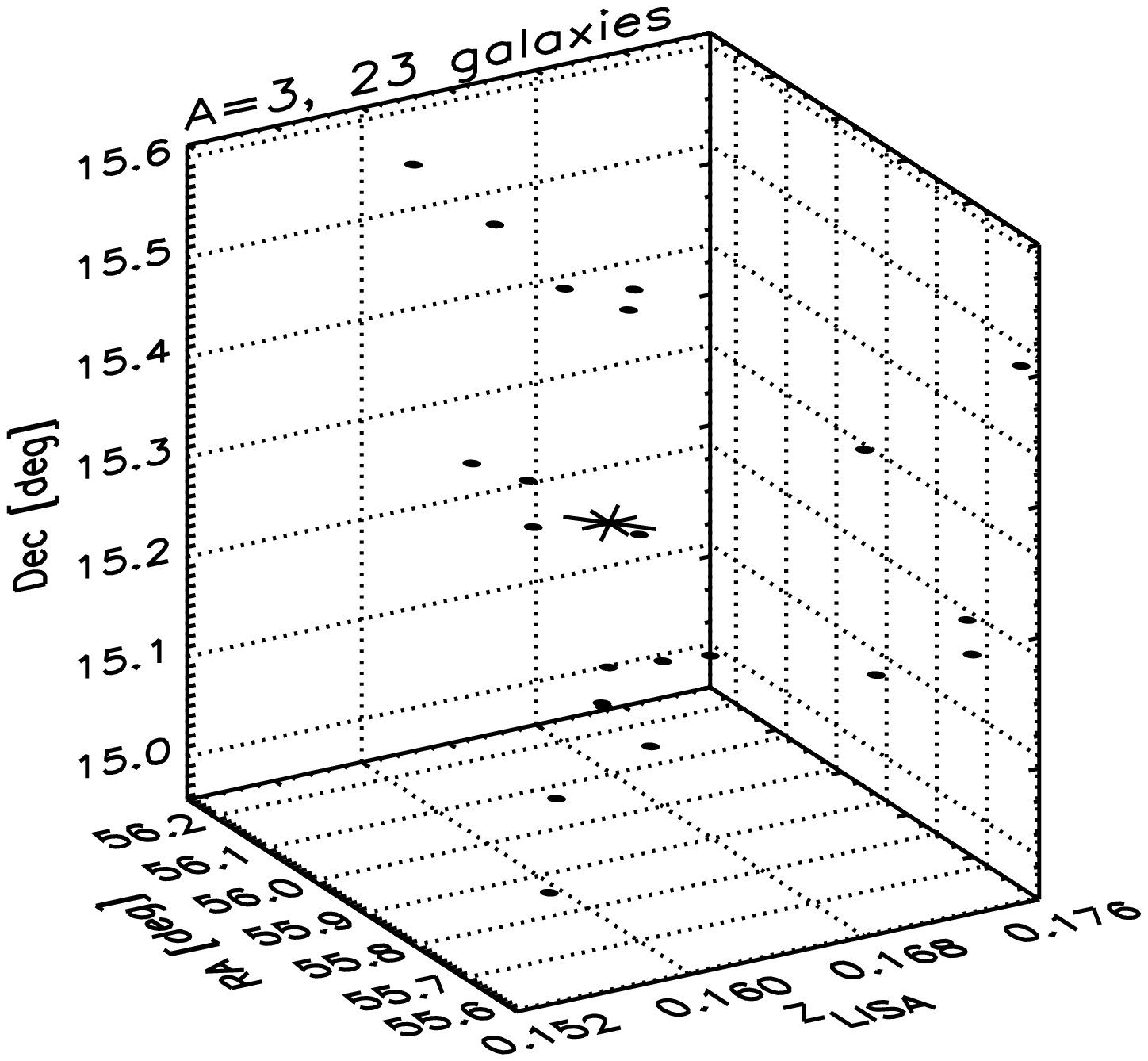}
     \includegraphics[width=2.2in]{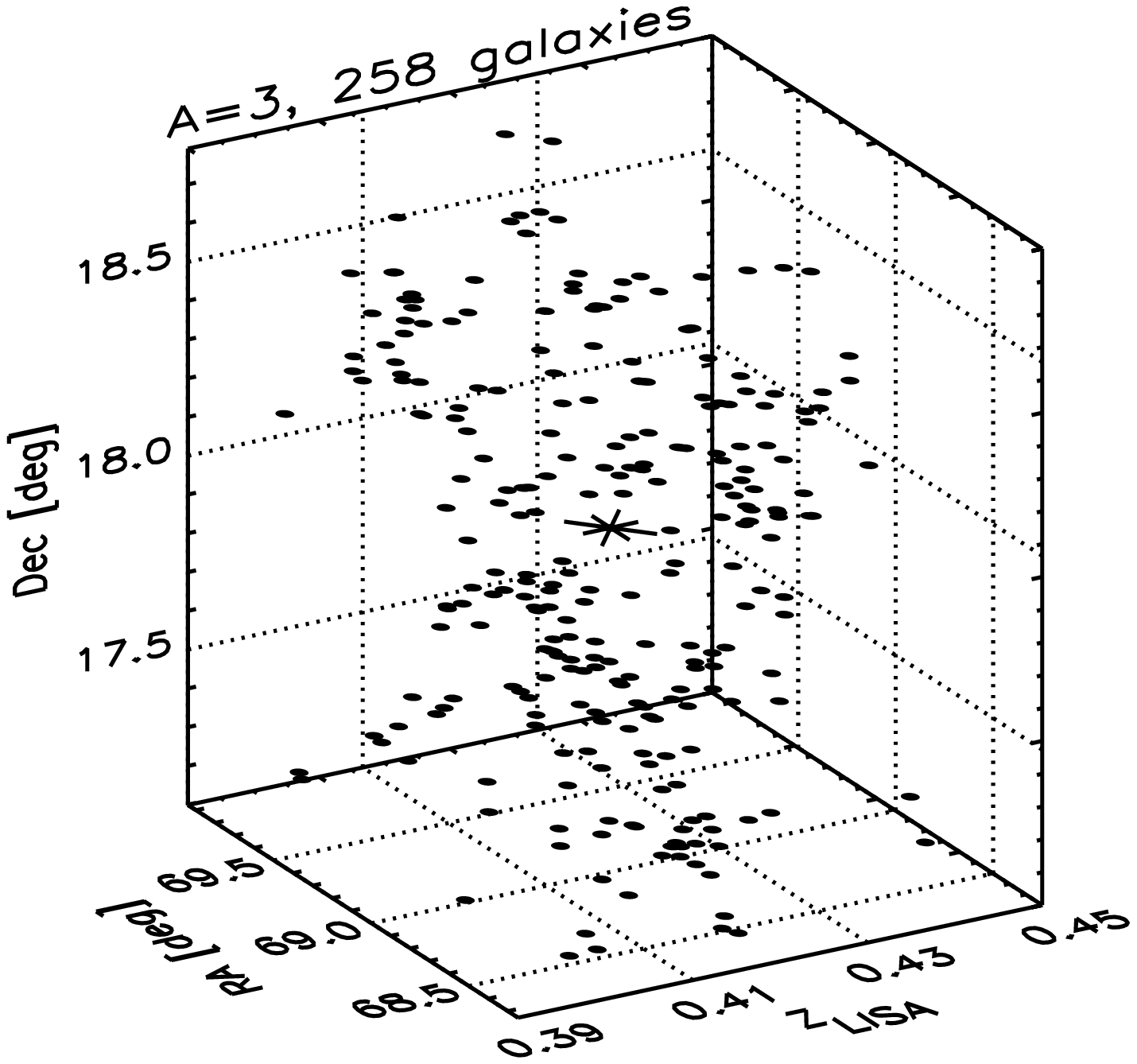}
   }\centerline{
     \includegraphics[width=2.2in]{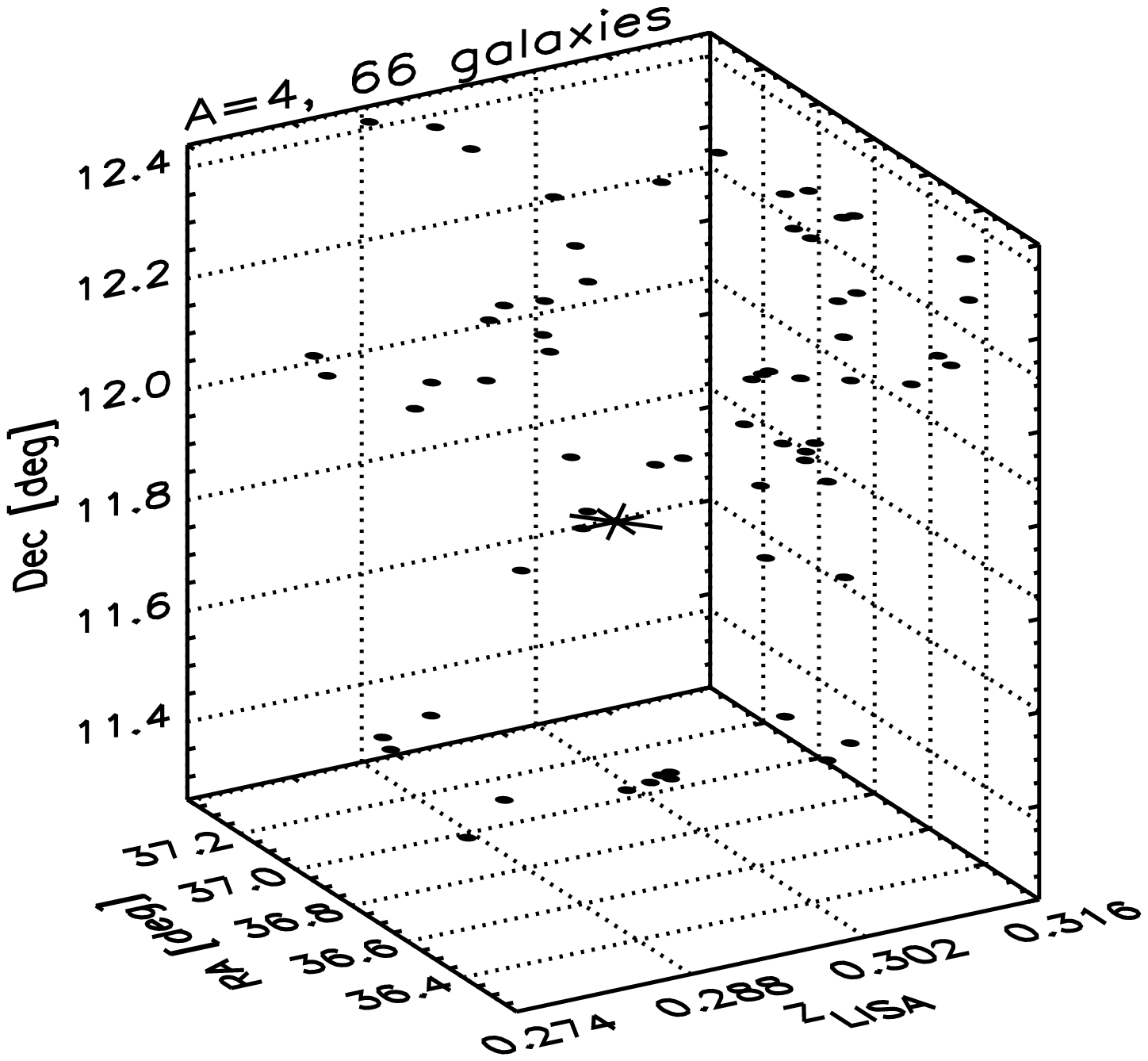}
     \includegraphics[width=2.2in]{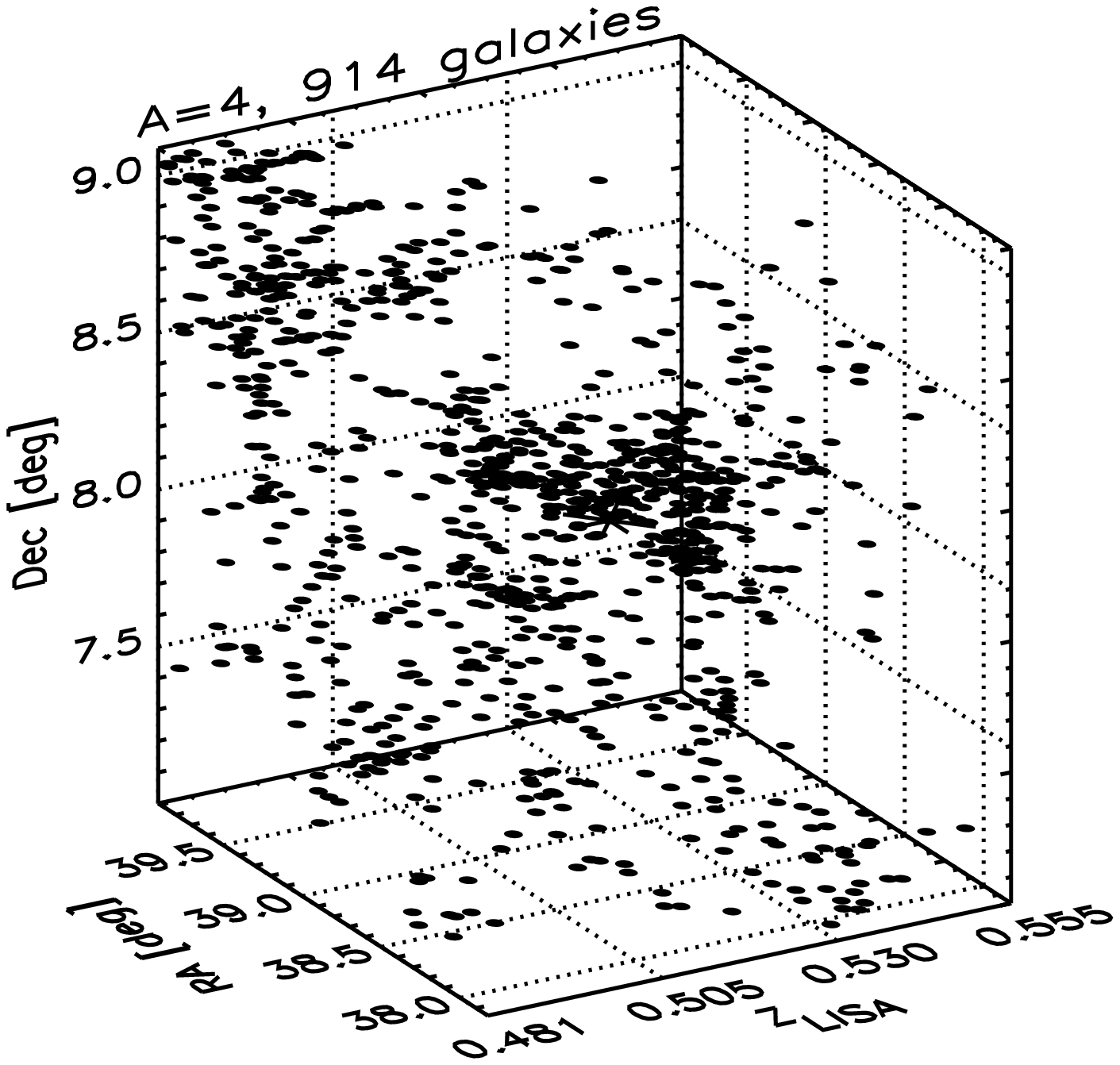}
   }\centerline{
     \includegraphics[width=2.2in]{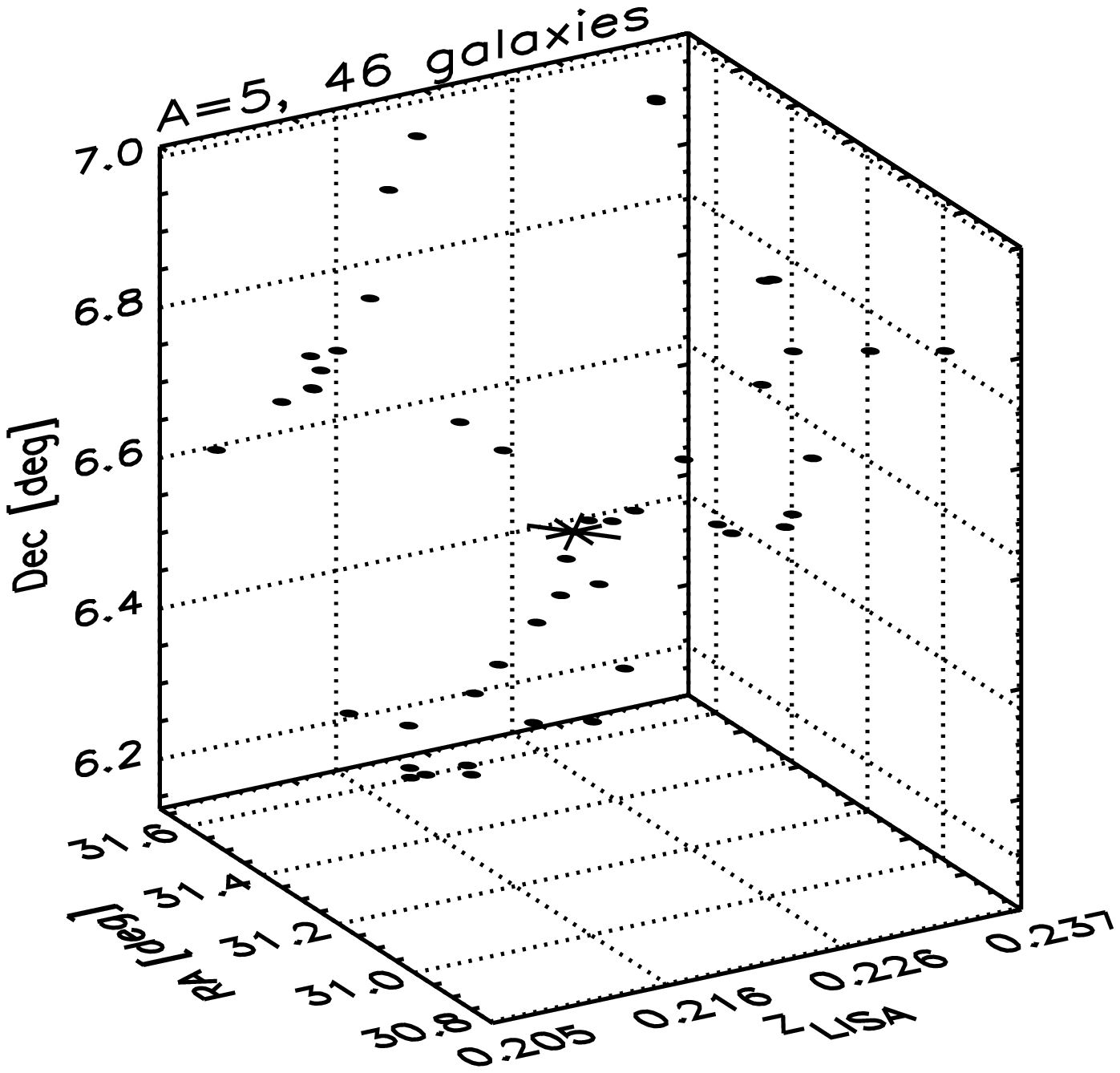}
     \includegraphics[width=2.2in]{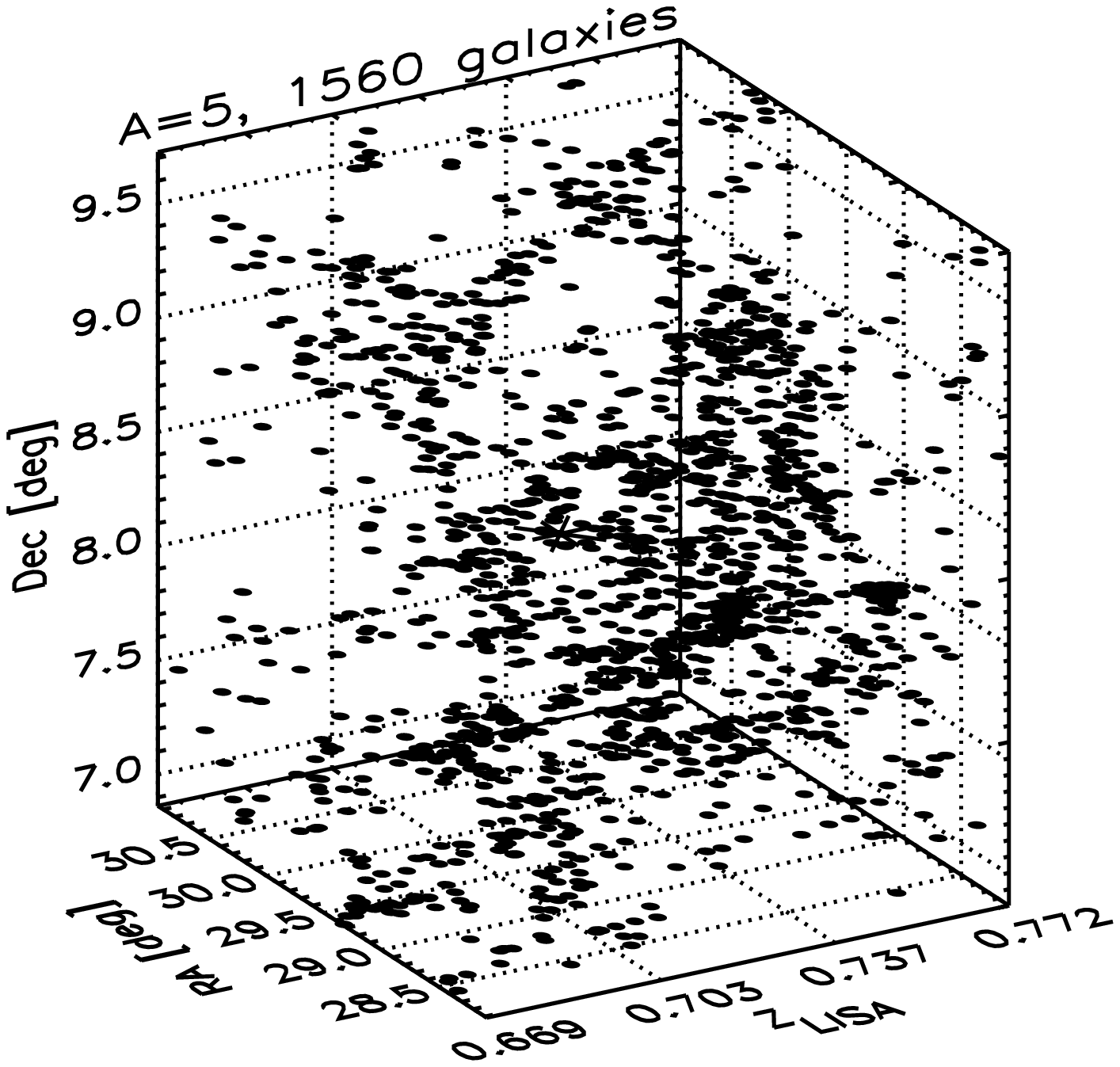}
   }
   \caption{Simulated LISA error boxes for one synthetic
   interferometer.  In each box the chosen source galaxy lies at the
   center, indicated by an asterisk, and dots represent all other
   galaxies in the box. Boxes are scaled to represent events at
   $z_{LISA} = Az_{SDSS}$, where $A$ is specified above each box, and
   the pretend LISA angular coordinates, given by $\theta_{LISA} =
   (1/A)\theta_{SDSS}$, are plotted.}
\label{fig:boxes}
\end{figure*}

To generate a mock EMRI catalog, a galaxy is selected at random in the
SDSS catalog and an error box is generated centered at this redshift
and direction.  Examples of simulated error boxes for the single
synthetic interferometer case are shown in
Fig.~\ref{fig:boxes}. While the ranges in right ascension and
declination are calculated using Cutler's estimates above, the range
in redshift corresponds to an error in $H_0$ of about 7\%
(representing a cosmological prior obtained from other techniques).
A histogram is then generated from
the SDSS galaxies in this box, with the originally selected host
galaxy removed. (Thus we assume that the actual host may not even be a
visible galaxy, only that it correlates with other galaxies in the
usual way.)   This gives the likelihood distribution  for the
EMRI host redshift and, according to (1), gives
$\ln{\cal L}_j(H_0)$ for this event.  Assuming a value $H_0 =$~70~km~s$^{-1}$~Mpc$^{-1}$, a linear Hubble's Law, and the same distance for all galaxies in
the box (calculated from the host redshift), we convert redshifts into
Hubble units. Thus our
histograms display a likelihood distribution for $H_0$ estimated for a sample, which should be compared with a ``true'' value of 70~km~s$^{-1}$~Mpc$^{-1}$.   

To simulate the LISA distance error for each source, each individual histogram is
displaced by a random amount, selected from an interval corresponding to
an error of $\Delta H_0 / H_0 = 0.05 z$ for two synthetic
interferometers, and $\Delta H_0 / H_0 = 0.07 z$ for one synthetic
interferometer.  As an example for the one interferometer case, given
a redshift of 0.2 the error would be $\Delta H_0 = 0.98$~km~s$^{-1}$~Mpc$^{-1}$, and the
histogram (that is, each ``box'') is displaced by a random amount between -0.49 and +0.49 in
$H_0$.  This error is applied before establishing the edges of
the histogram at $H_0=65$ and $H_0=75$~km~s$^{-1}$~Mpc$^{-1}$, so that new galaxies are
allowed to enter the histogram due to the shift in $H_0$. 

A  set of such samples can be combined by stacking their histograms
(appropriately normalized to give all events equal weight independent
of the number of galaxies in their sample, as in (1)).  Since each box is generated
with the originally chosen source galaxy assuming $H_0 =
$~70~km~s$^{-1}$~Mpc$^{-1}$, a peak should emerge around this value
in the stacked histogram.  The
width of the summed distribution measures the width of the
distribution of galaxies, and the offset of the fitted mean from the
center (70~km~s$^{-1}$~Mpc$^{-1}$) measures the actual offset in the ``measured'' from the ``true'' value of $H_0$ for each set of realizations.

\begin{figure*}[htbp]
   \centerline{
     \includegraphics[width=2.5in]{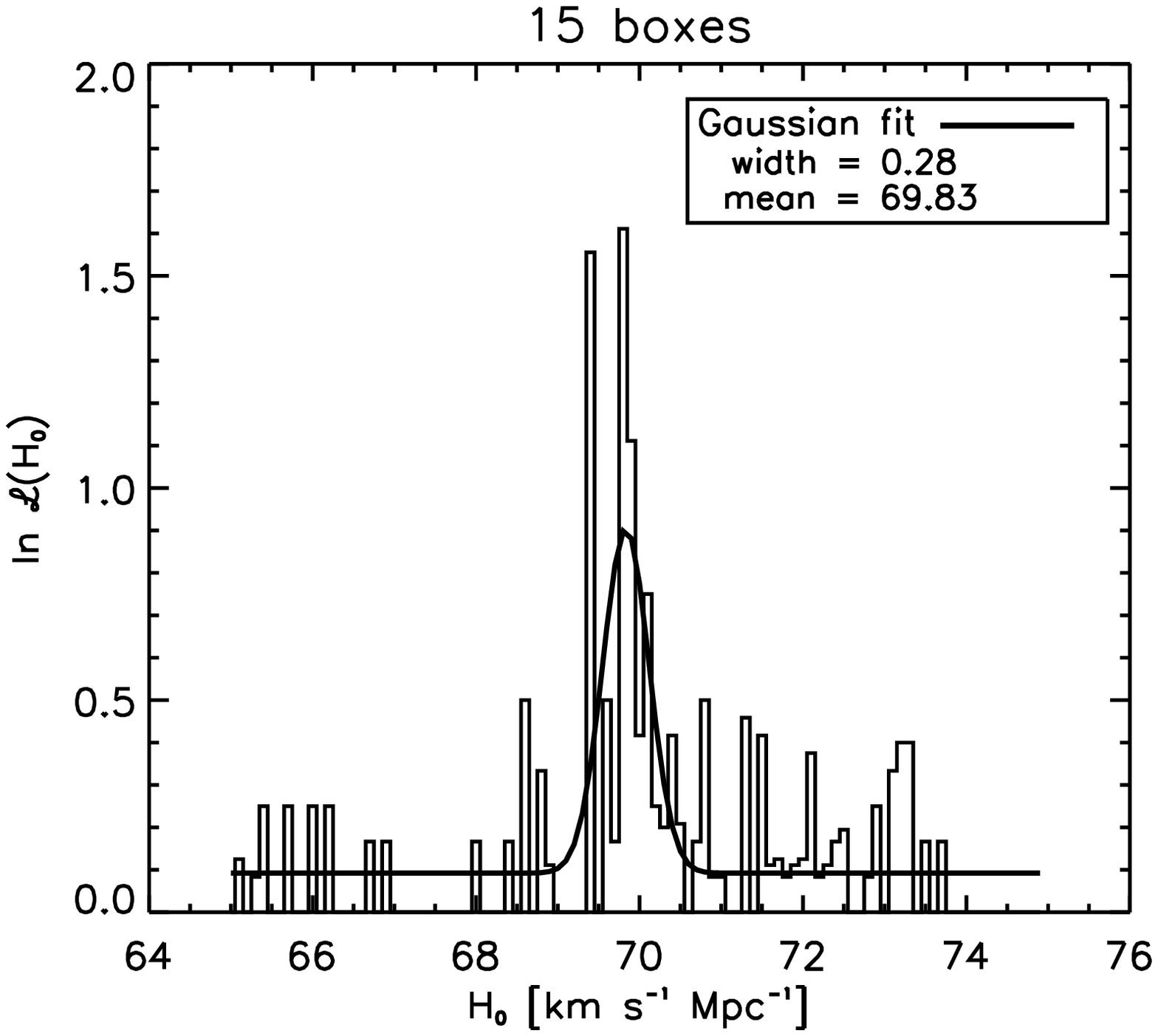}
     \includegraphics[width=2.5in]{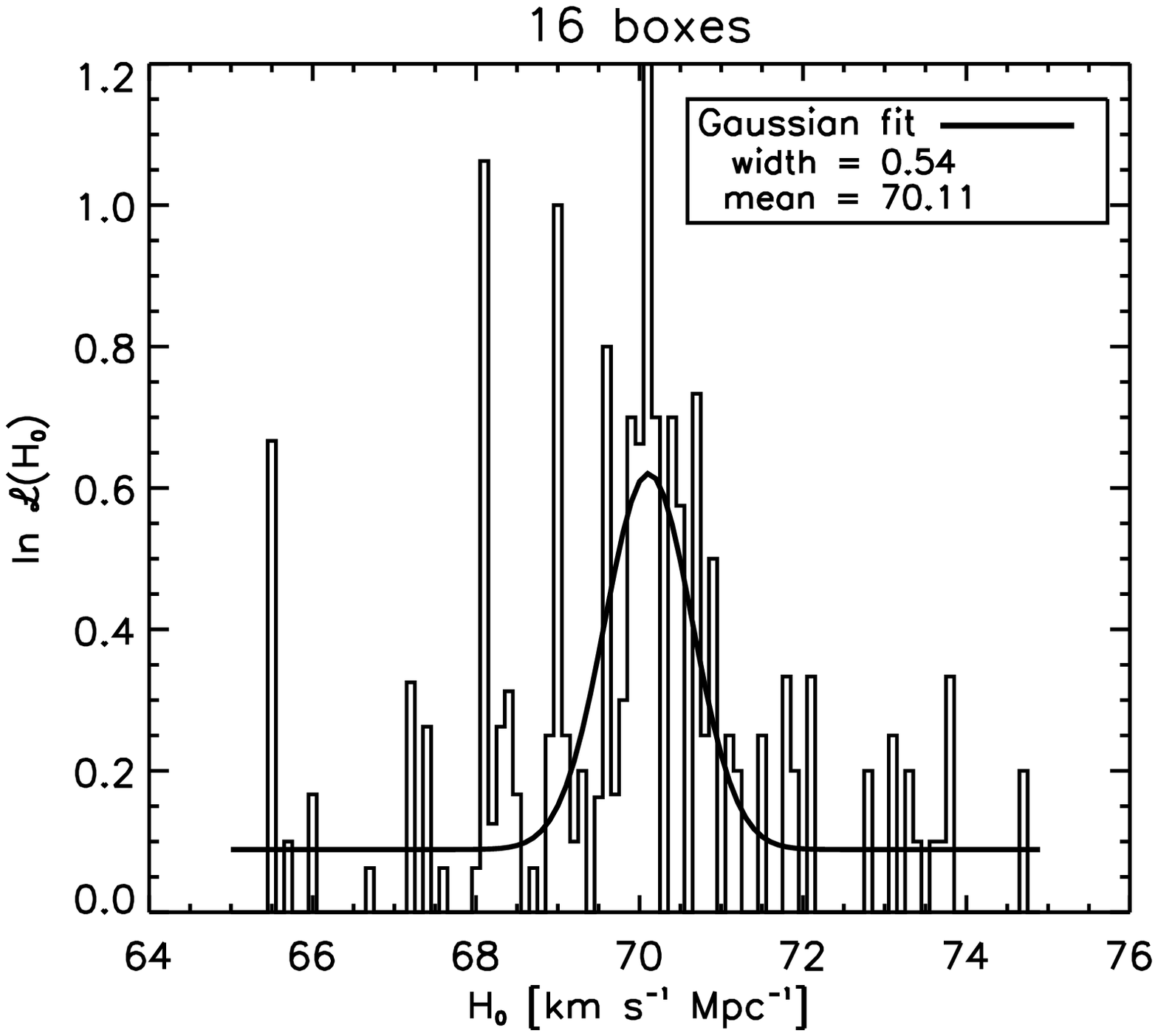}
     \includegraphics[width=2.5in]{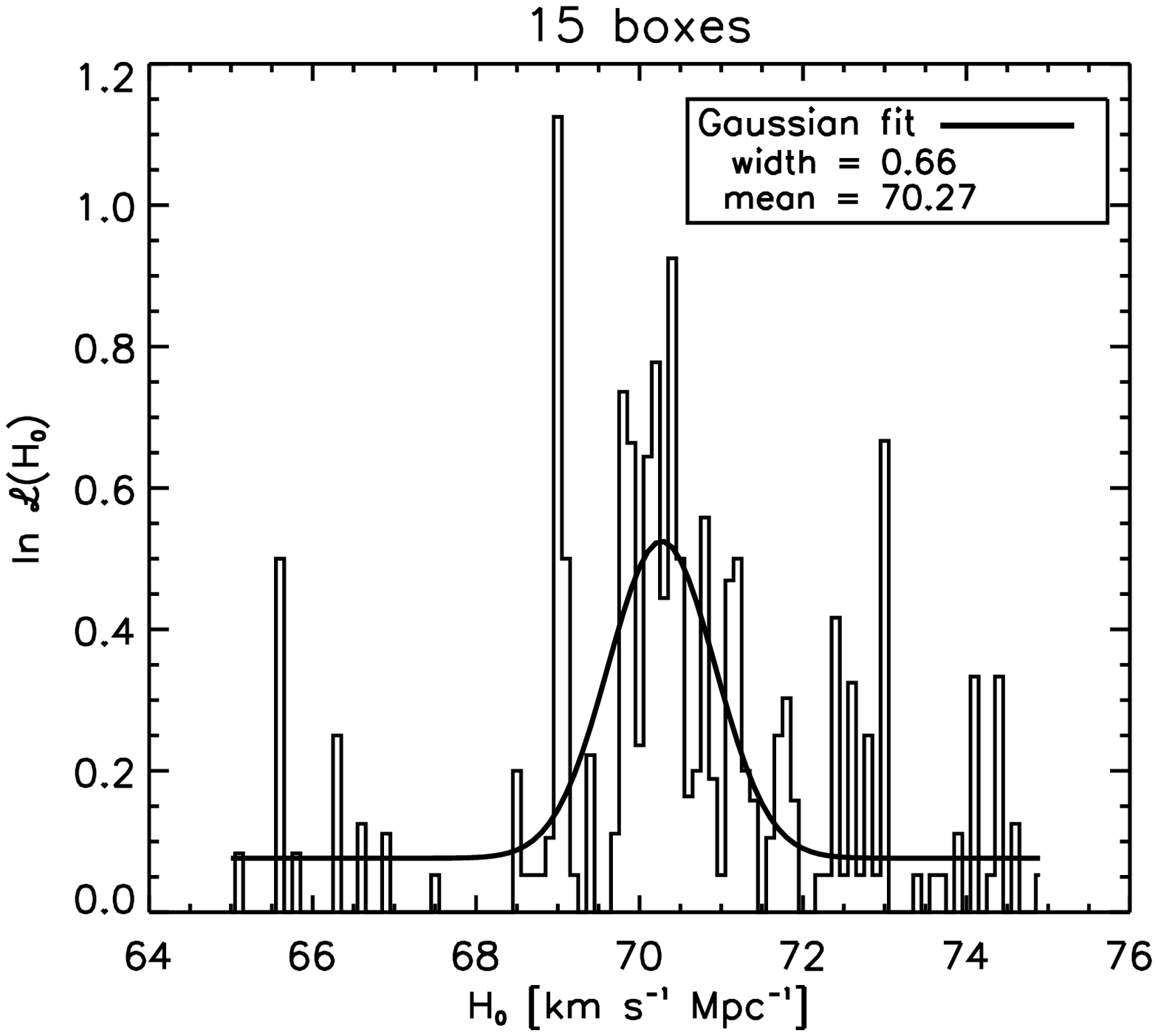}
   }
   \centerline{
     \includegraphics[width=2.5in]{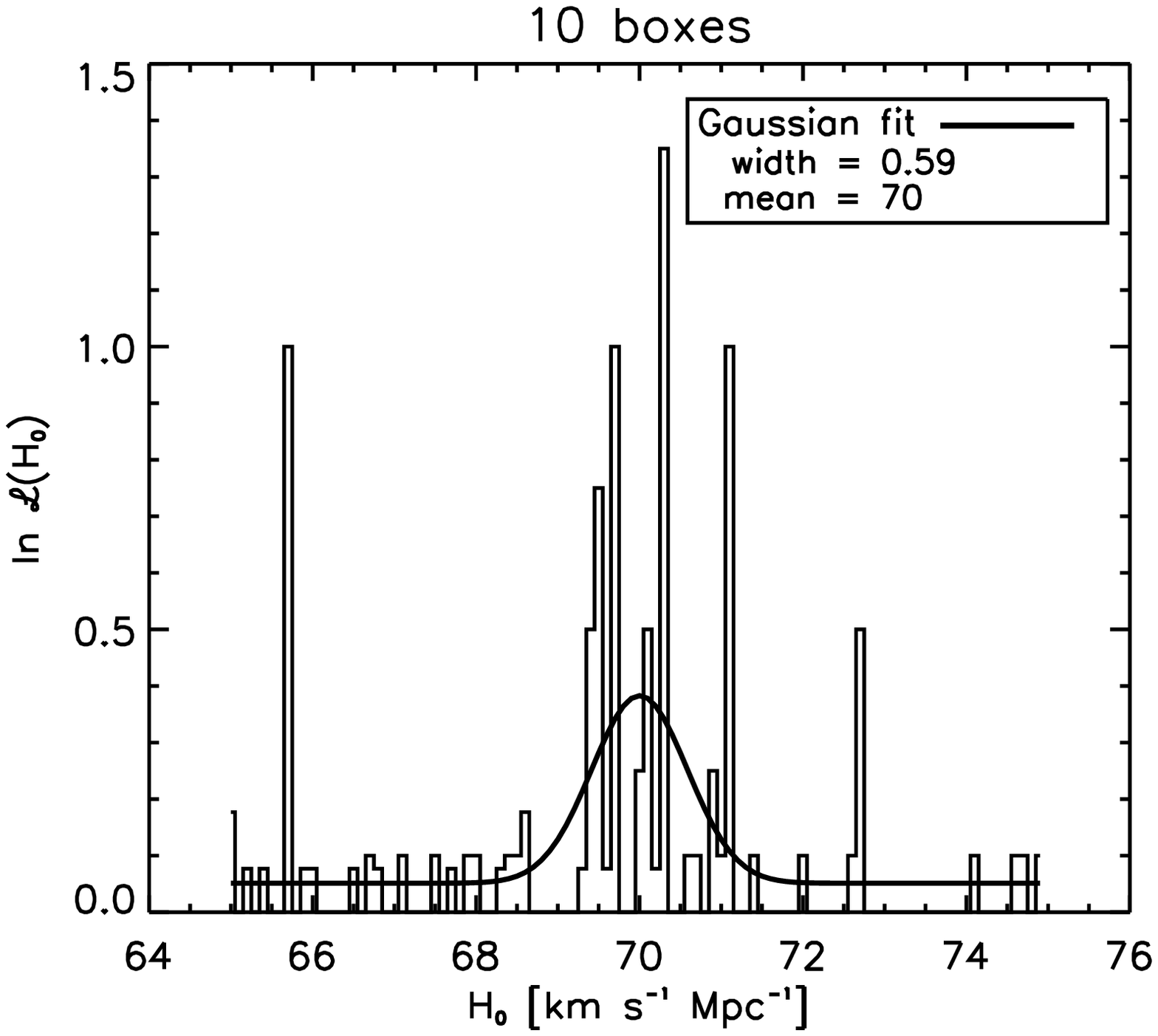}
     \includegraphics[width=2.5in]{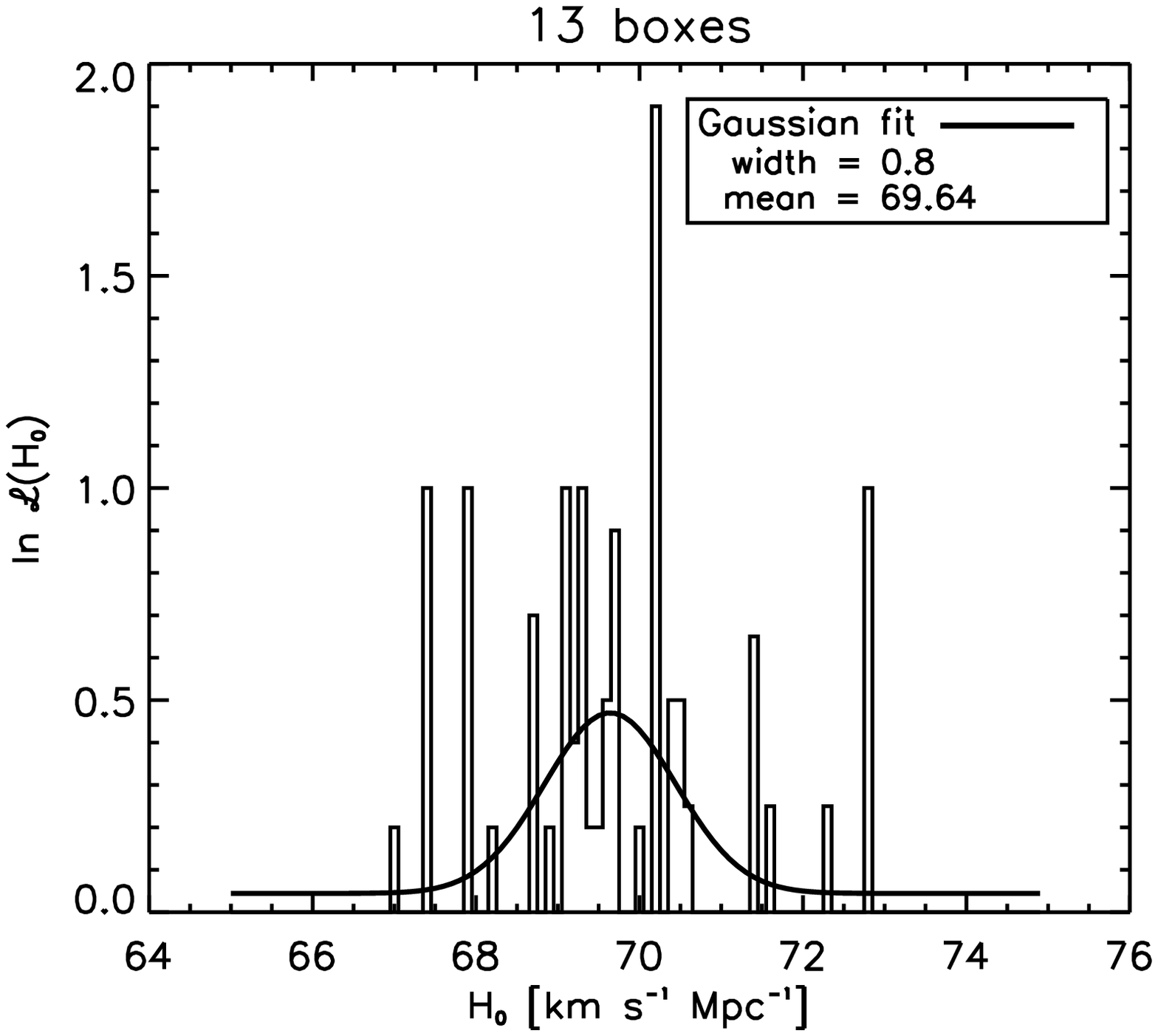}
     \includegraphics[width=2.5in]{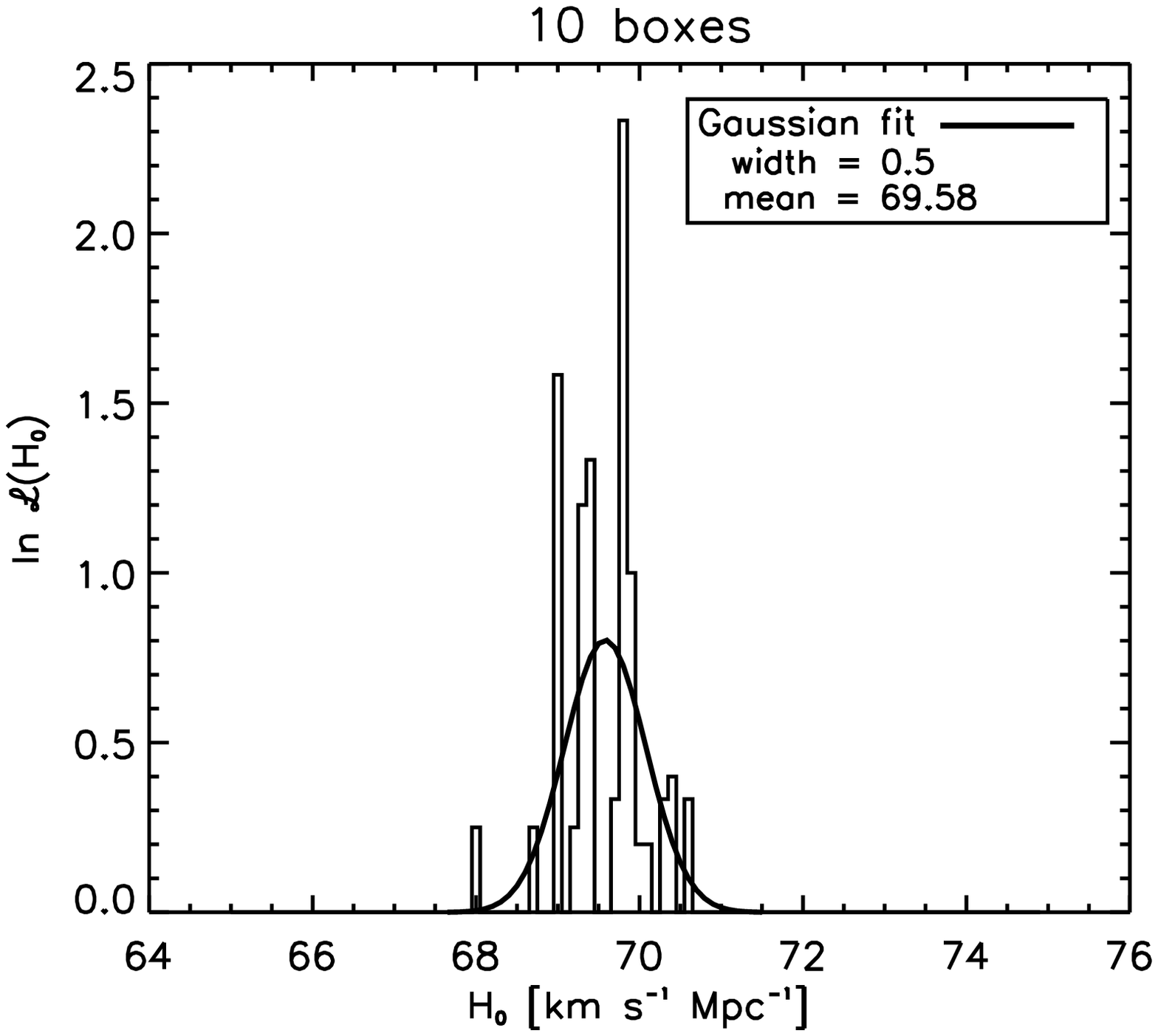}
   }
   \caption{Summed histograms (bin size = 0.1) with
   Gaussian fits overlayed.  Top row: one synthetic
   interferometer case; bottom row: two synthetic interferometers.
   The true value of $H_0$ was assumed to be 70~km~s$^{-1}$~Mpc$^{-1}$. 
   Gaussian fits were generated using a Levenberg-Marquardt technique
   to iteratively search for a best least-squares fit to $\ln{\cal
   L}(H_0)$.
   The starting value for the mean was a random value between 68 and
   72, and those for the width and area were both 1. Vertical axes are in arbitrary units.}
\label{fig:A1}
\end{figure*}

Results of several mock realizations are displayed in Fig.~\ref{fig:A1}. For each of
 these sums, 20 ``EMRIs'' were randomly selected in the SDSS (DR6)
 spectroscopic survey
 volume (673,264 galaxies total, classified spectroscopically), excluding the distant part of the survey where
 galaxy selection is dominated by different criteria. The
 (spectroscopic) redshifts were limited to
 $0.02 \leq z \leq 0.23$, and all mock error boxes were limited to the regions of the northern galactic
 cap covered by DR6 (shown on the SDSS website, which is listed under `Acknowledgements'). The SDSS provides a statistically complete sample for galaxies with r-band
Petrosian magnitudes $r \leq 17.77$ \cite{2002AJ....124.1810S}. In our analysis we neglect peculiar velocities of galaxies, which  
are likely to add errors of less than about 1\% per object even  
for the nearest plausible samples and are therefore subdominant. 

Each individual error box that was
 constructed yields $\ln{\cal L}_j(H_0)$, displayed as a histogram which plots the number of galaxies versus
 $H_0$.  The originally chosen source galaxy was subtracted from each
 histogram, and therefore error boxes that contained only the
 chosen source galaxy were ignored.  This is reflected in the final
 box count (the initial box count is 20), which is listed above each
 plot. Each of these histograms were then normalized by dividing
 by the total number of galaxies contained in each error box
 ($N_j$), not including the original source galaxy. Individual
 histograms were then added together, resulting in a plot of $\ln{\cal L}(H_0)$ that contains information purely from galaxy clustering.

The results from these realizations are summarized in Table~\ref{tab:A1}, where
the average redshift of the chosen source galaxies is included, as well as the total number
of galaxies used in the summed histogram. It can
be seen that the samples cluster near the ``true'' value of 70~km~s$^{-1}$~Mpc$^{-1}$
and would allow a measurement of $H_0$ with a typical error in the
mean of about
$0.2$~km~s$^{-1}$~Mpc$^{-1}$ (averaged over 25 realizations for
the one interferometer case), or a precision of about one third of a percent. This shows that the
statistical redshift  technique is credible  with a realistic galaxy
distribution and LISA errors.

 As expected,  the distribution of  
candidate Hubble constants is highly nongaussian, so a sample at least this large is needed
 for a reliable result. In order to get a sample of order 10 or more
 EMRIs at such a low redshift, the overall rate of EMRIs would have to
 be at the high end of current estimates \cite{2006ApJ...645L.133H}.

\begin{table*}
\caption{Results for realizations in Fig.~\ref{fig:A1}; starting from
  top left.}
\begin{tabular}{c c c c c}
\hline\hline
Number of Boxes & Average $z$ & $\Sigma N_j$ & Width of Gaussian
& Mean $H_0$ \\
 (Sources)& &  &
& [km~s$^{-1}$~Mpc$^{-1}$]\\
\hline
15  & 0.125  & 72 & 0.28  & 69.83\\
16  & 0.109  & 74 & 0.54  & 70.11\\
15  & 0.135  & 103 & 0.66  & 70.27\\
10  & 0.119  & 46 & 0.59  & 70.00\\
13  & 0.113  & 32 & 0.80  & 69.64\\
10  & 0.114  & 21 & 0.50  & 69.58\\
\hline\hline
\end{tabular}
\label{tab:A1}
\end{table*}

Therefore, in a second round of experiments, we tested the reliability of the
above technique for larger redshifts, still using the same SDSS data (with $z$ limited to $0.02 \leq z \leq 0.23$)
but scaling to estimate higher redshift behavior in the same galaxy
population. We chose to scale to higher $z$ from this SDSS range
since the SDSS does not sample typical galaxy populations at $z \gtrsim 0.23$. 

Adopting a larger redshift value
for a fiducial EMRI event increases the LISA angular and distance errors according to
Cutler's estimates, and at the same time requires a scaling of the SDSS data in order to estimate the properties of a similarly-clustered galaxy
host population at larger redshift.
For one synthetic interferometer, Cutler's estimates give us an
angular error $\Delta \theta_{LISA} = 4z_{LISA}$~degrees on each side of
an error box.  When scaling by a factor $A$ to a source redshift
$z_{LISA} = Az_{SDSS}$, where $z_{SDSS}$ is the redshift of a random host galaxy in
SDSS, the LISA angular error is 
\begin{equation}
\Delta \theta_{LISA} = 4Az_{SDSS} \rm ~degrees.
\end{equation}
To estimate the angle corresponding to the galaxy structure at higher
redshift in this angular box, we assume a
$\Lambda$CDM cosmology with a few simplifications.  Assume that the
galaxy distribution is frozen in comoving coordinates, which is approximately
true in a $\Lambda$ dominated universe.  Then a structure that has an
angular size $\theta_{LISA}$ at redshift $z_{LISA}$ will appear at
$z_{SDSS}$ with an angular size
\begin{equation}
\theta_{SDSS}= \theta_{LISA}  (D_{LISA}/D_{SDSS})[ (1+z_{LISA})/(1+z_{SDSS})  ]
\end{equation}
where angular size distances are denoted by $D$.  One can verify
numerically in $\Lambda$CDM that within 10\%, this agrees with a simple formula,
\begin{equation}
\theta_{SDSS}= \theta_{LISA} (z_{LISA}/z_{SDSS}),
\end{equation}
which can be obtained by using a small $z$ expansion and neglecting higher order
terms in $z$ (which is appropriate since the range in
$z$ in a LISA error box is small).  Therefore, combining (3) and (5), the angular size we use on one side of an SDSS error
box in our scaled realizations scales as
\begin{equation}
\Delta \theta_{SDSS} = \Delta \theta_{LISA} A = 4A^{2}z_{SDSS}  \rm ~degrees.
\end{equation}

\begin{figure*}[htbp]
   \centerline{
     \includegraphics[width=2.5in]{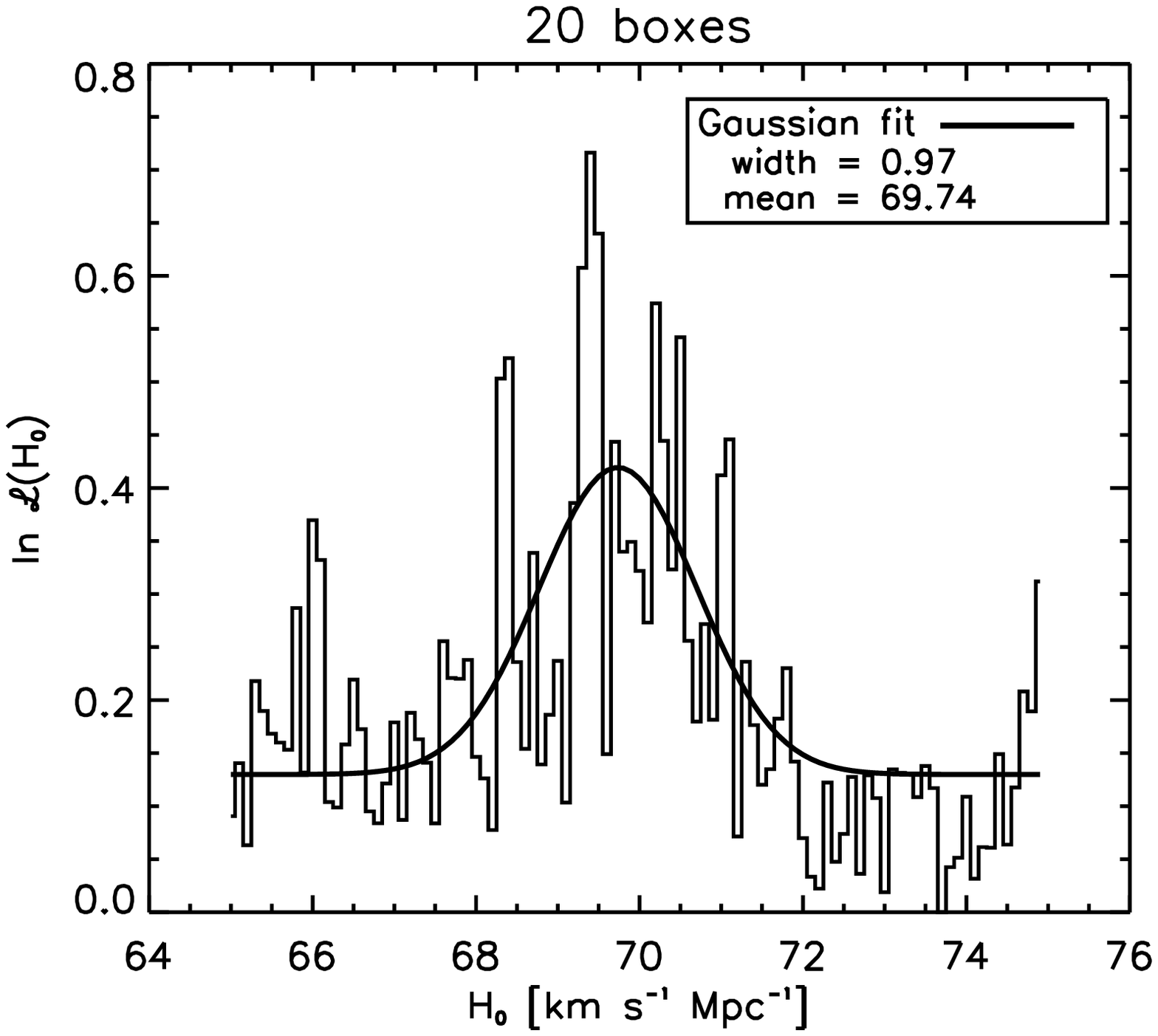}
     \includegraphics[width=2.5in]{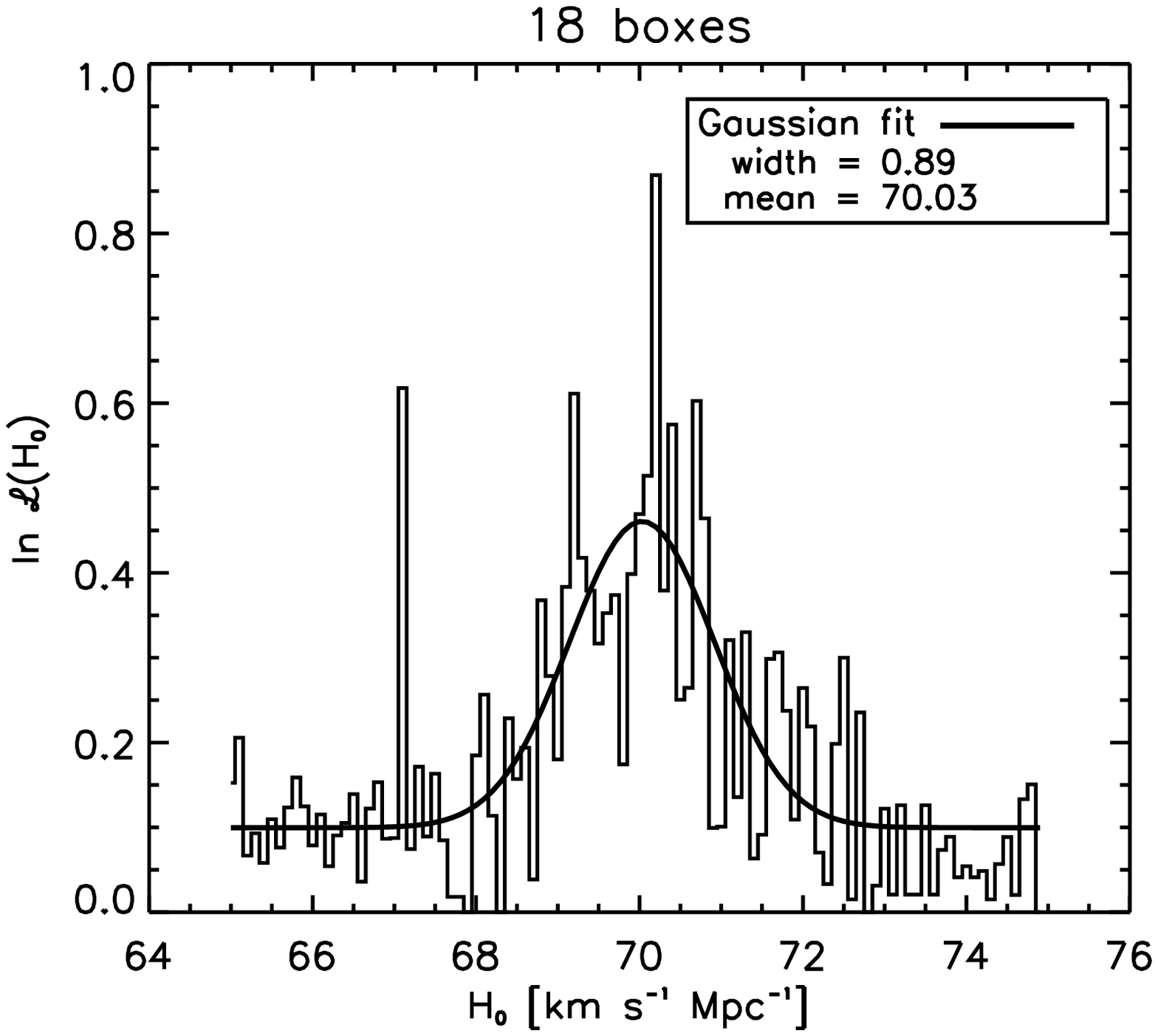}
     \includegraphics[width=2.5in]{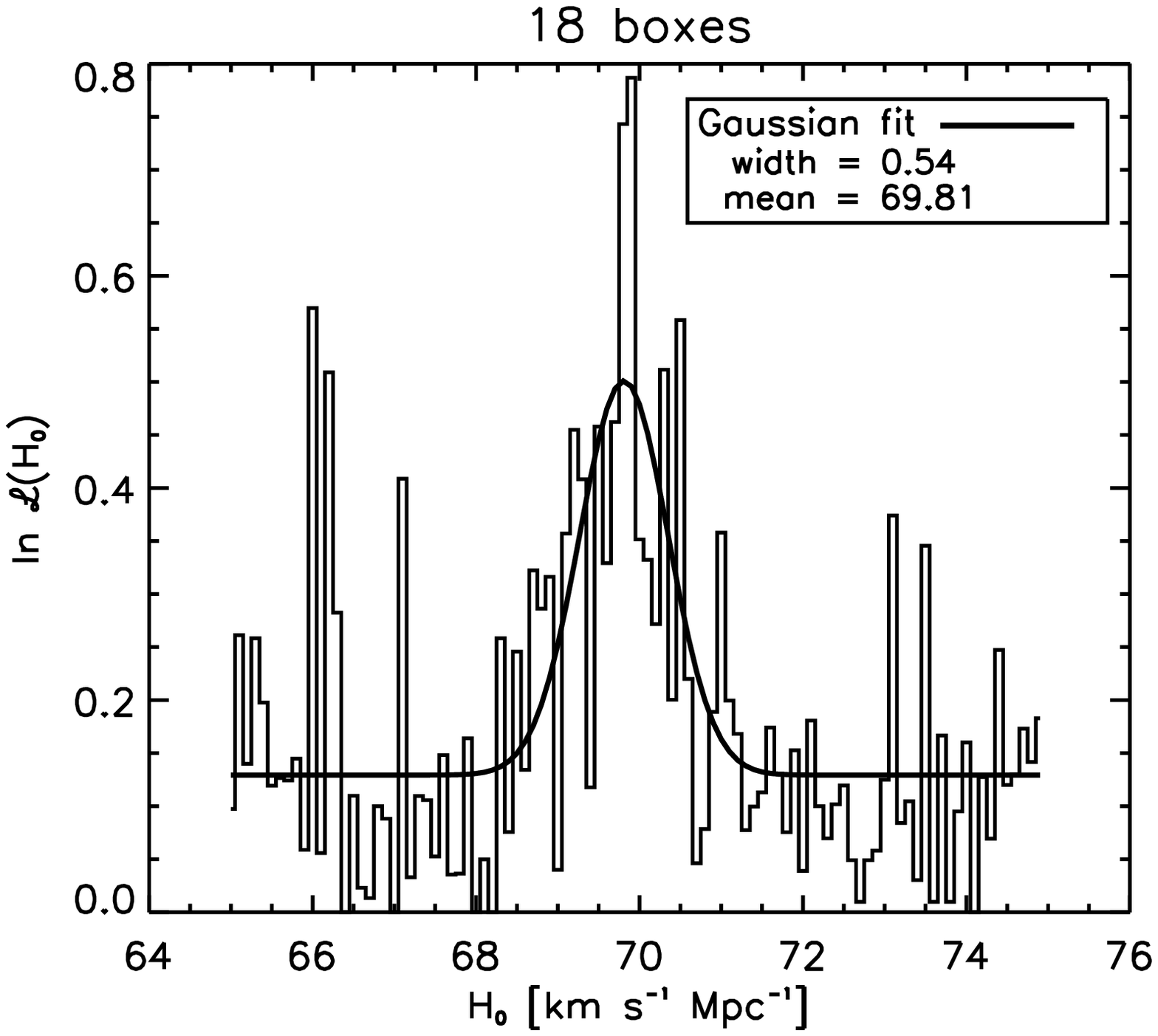}
   }\centerline{
     \includegraphics[width=2.5in]{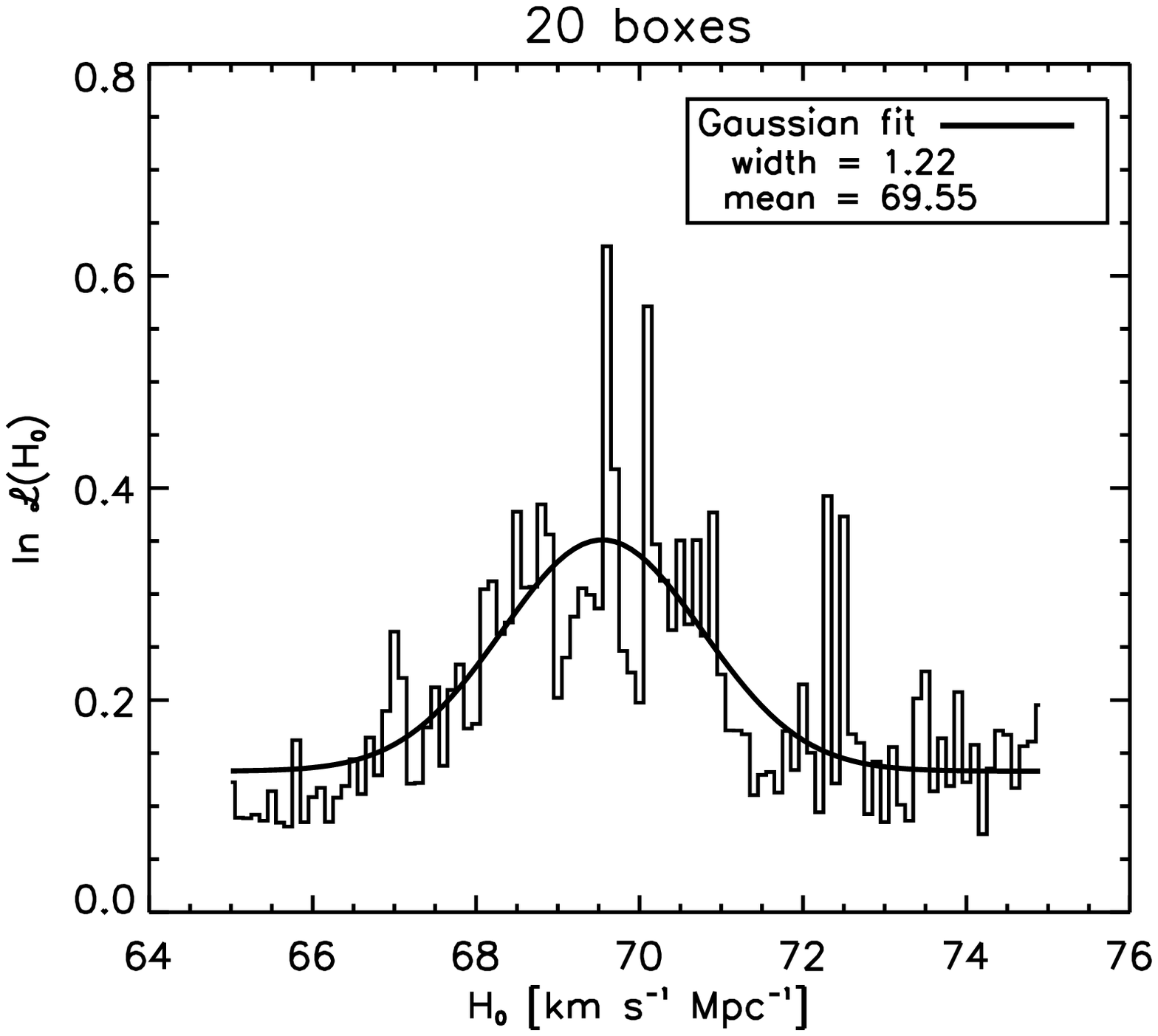}
     \includegraphics[width=2.5in]{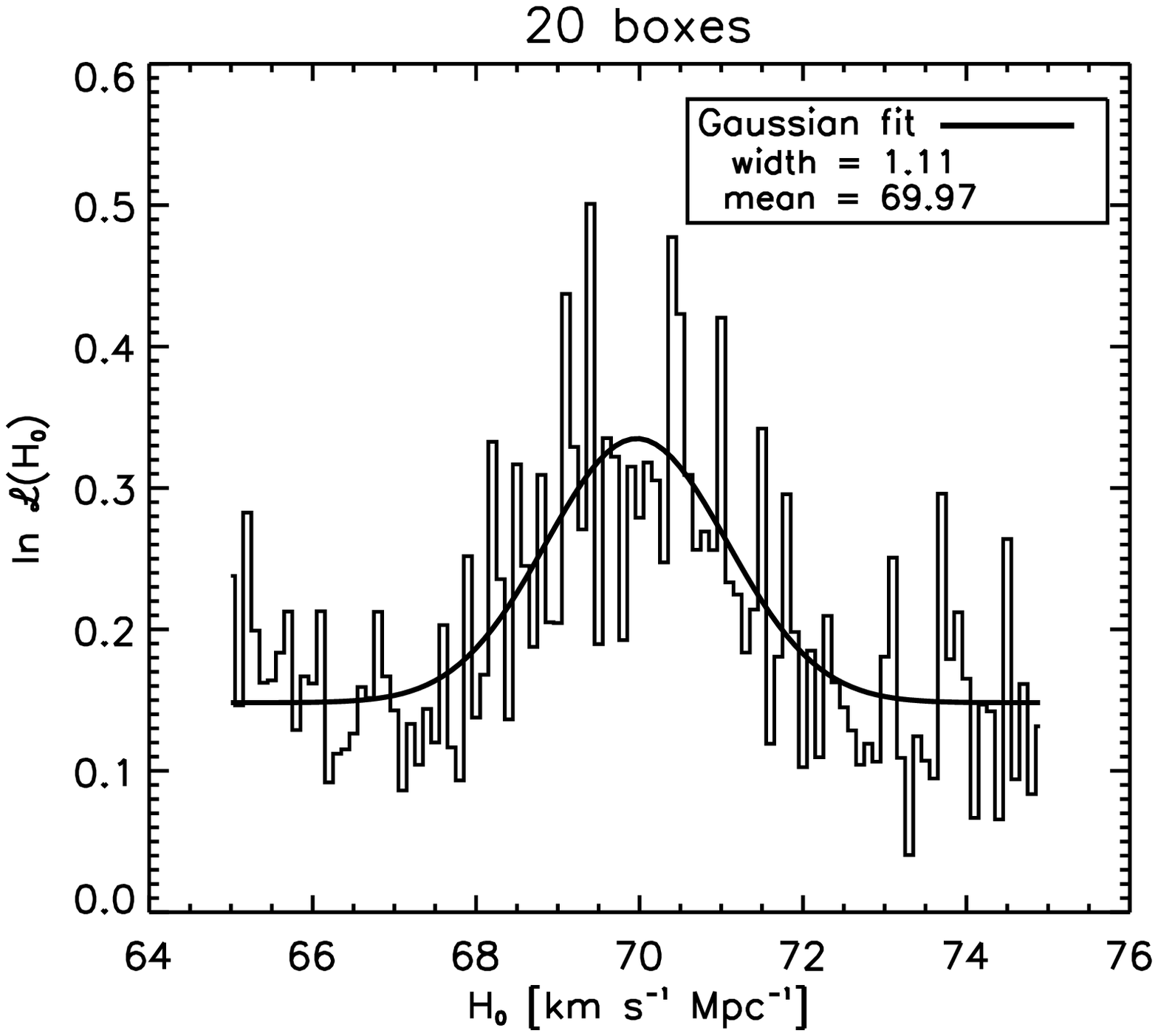}
     \includegraphics[width=2.5in]{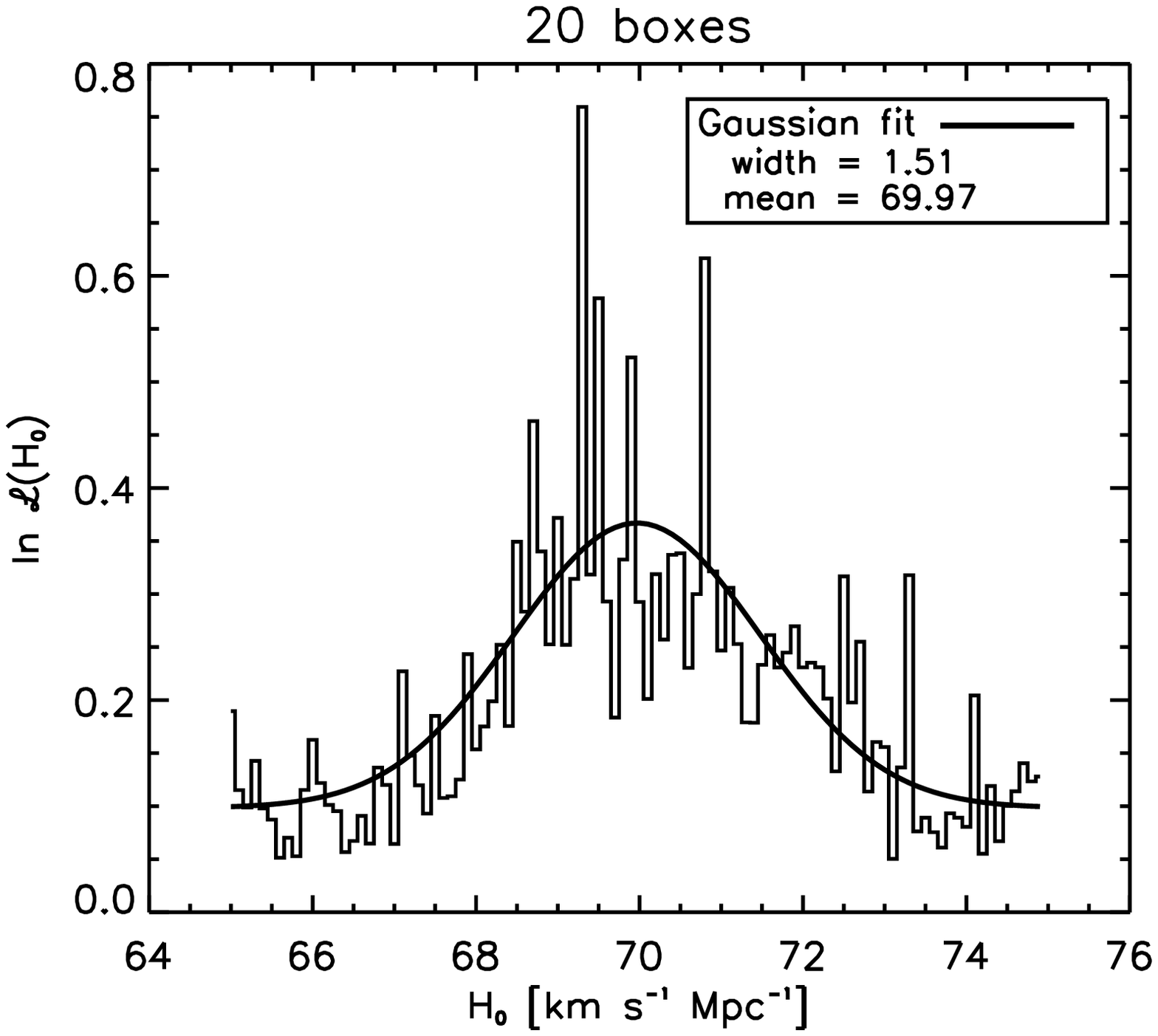}
   }\centerline{
     \includegraphics[width=2.5in]{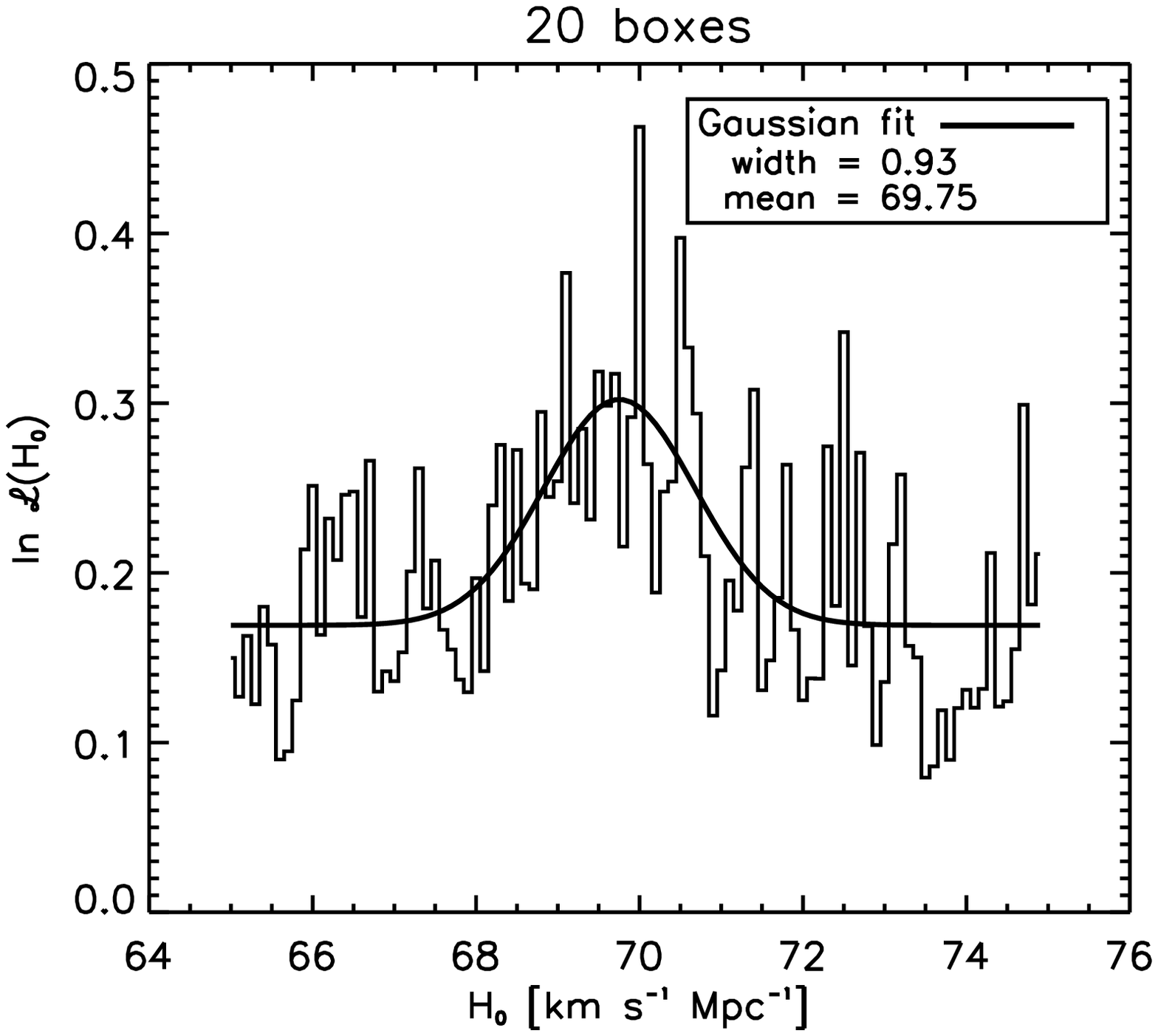}
     \includegraphics[width=2.5in]{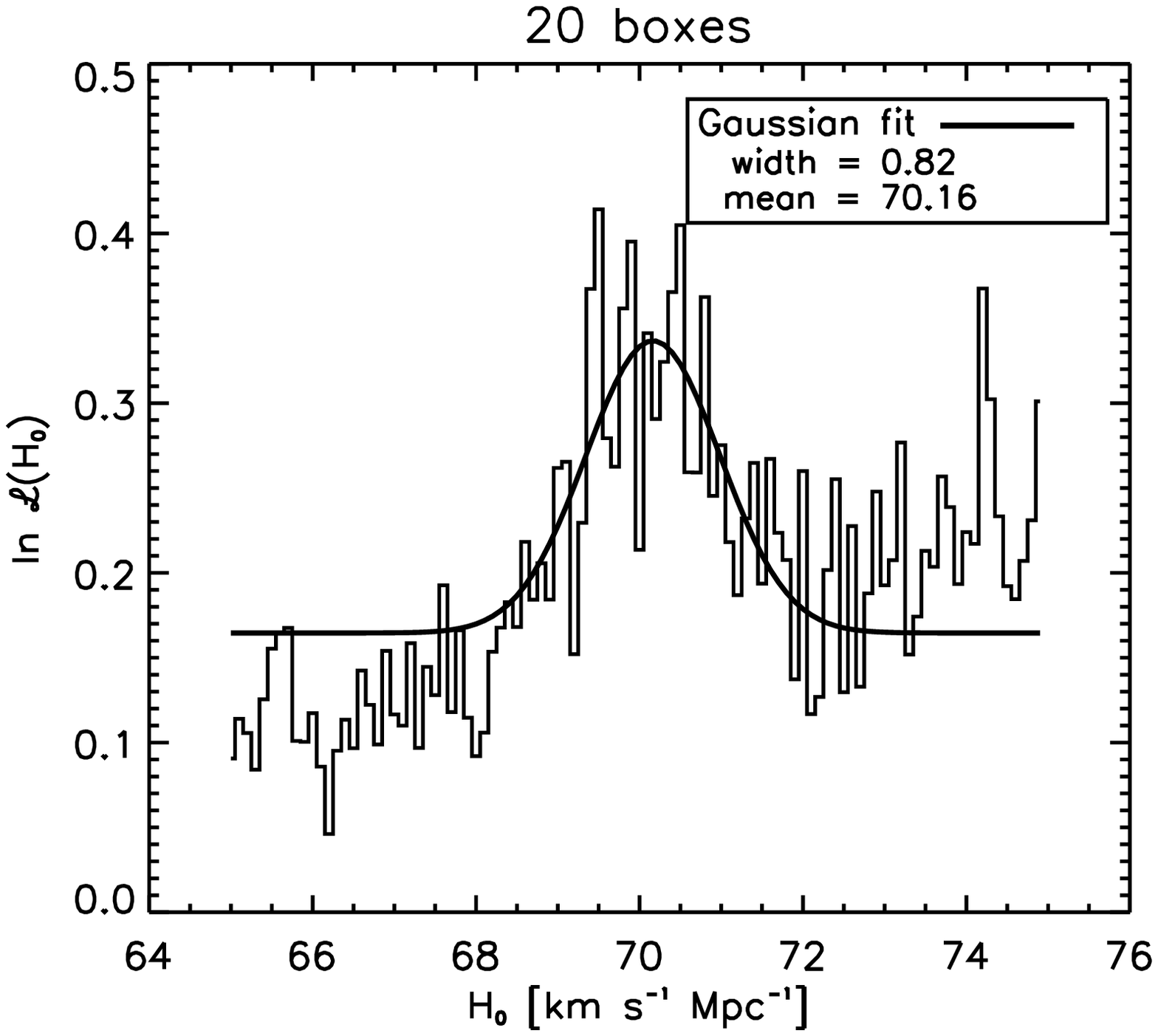}
     \includegraphics[width=2.5in]{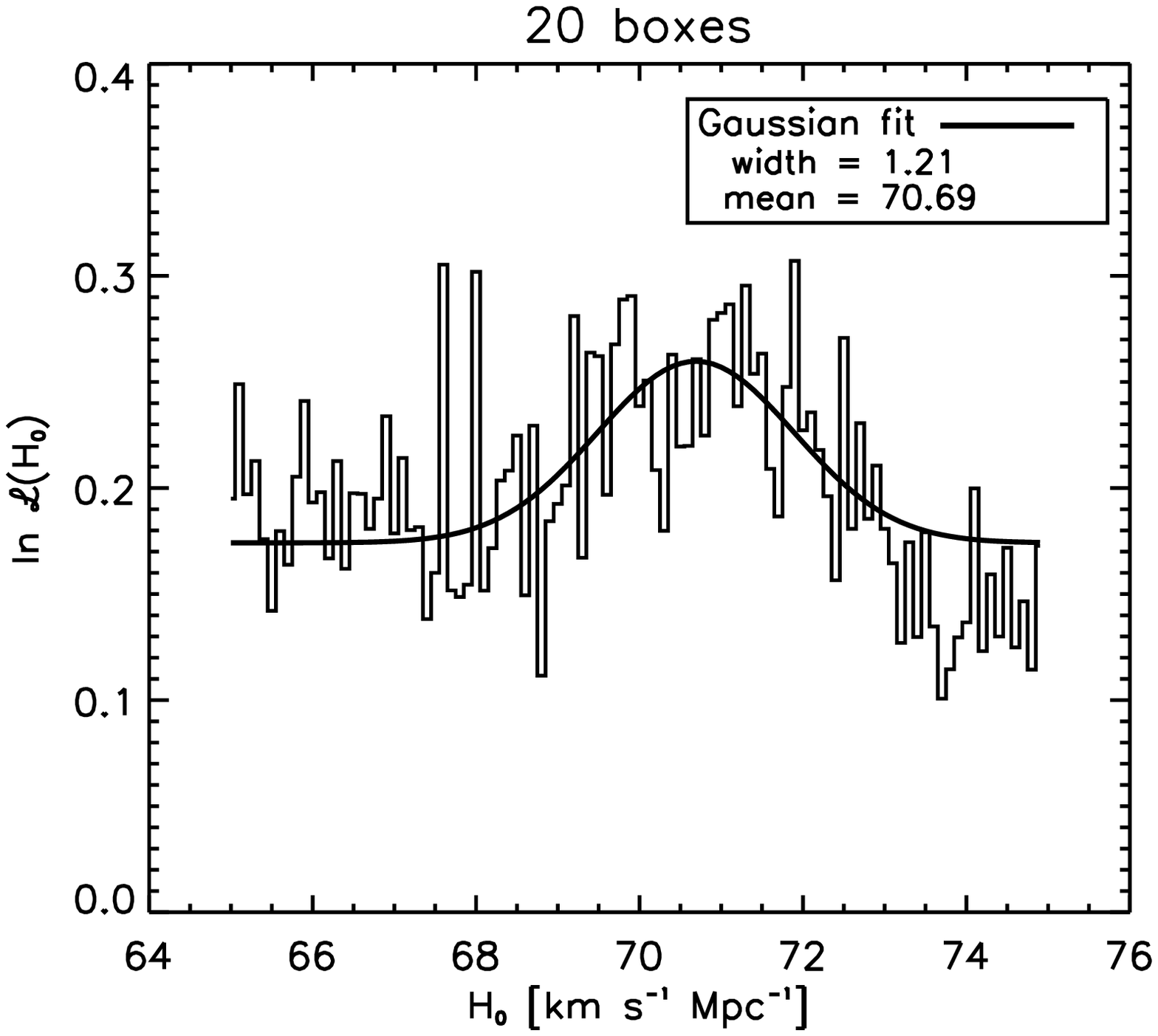}
   }\centerline{
     \includegraphics[width=2.5in]{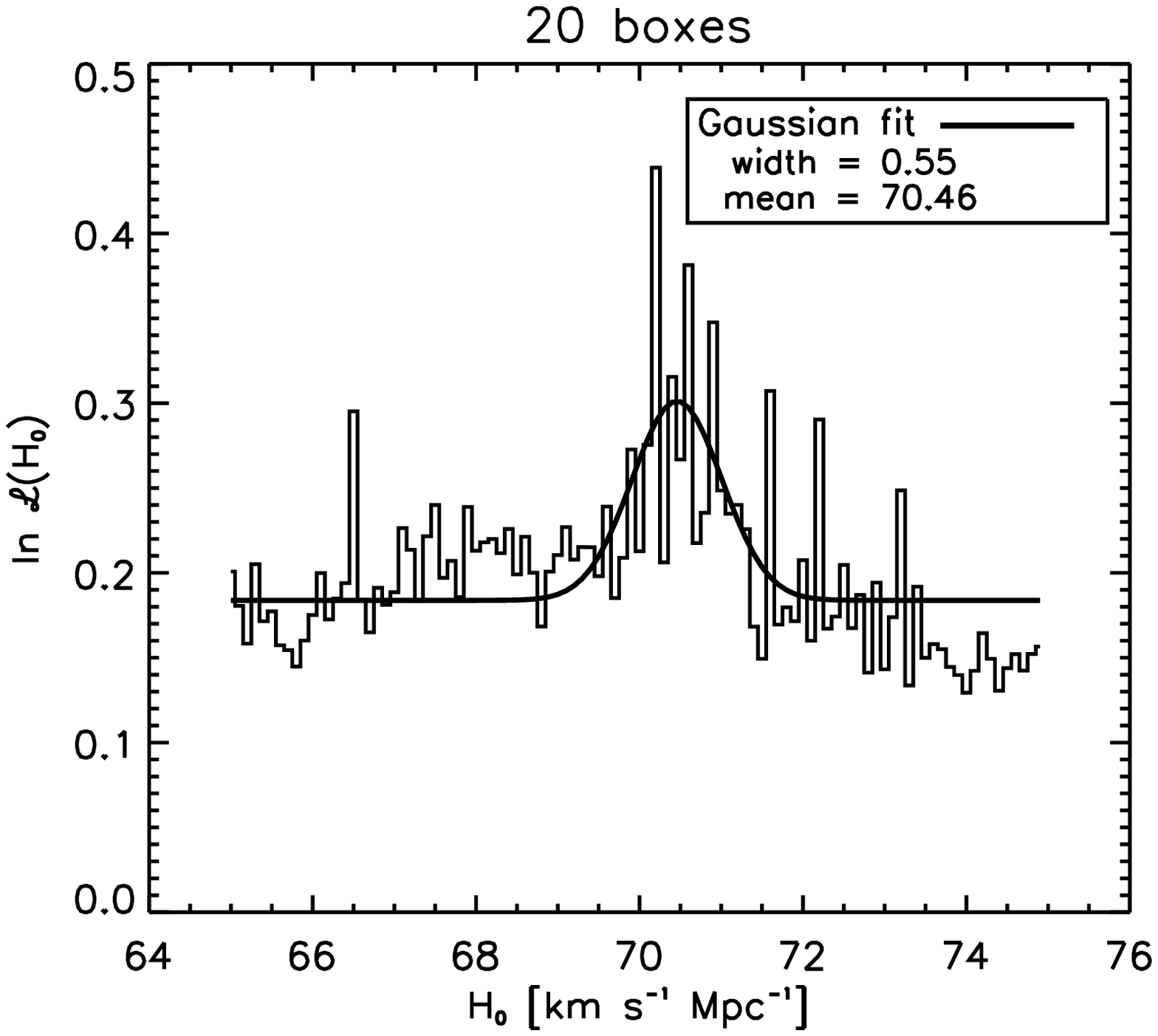}
     \includegraphics[width=2.5in]{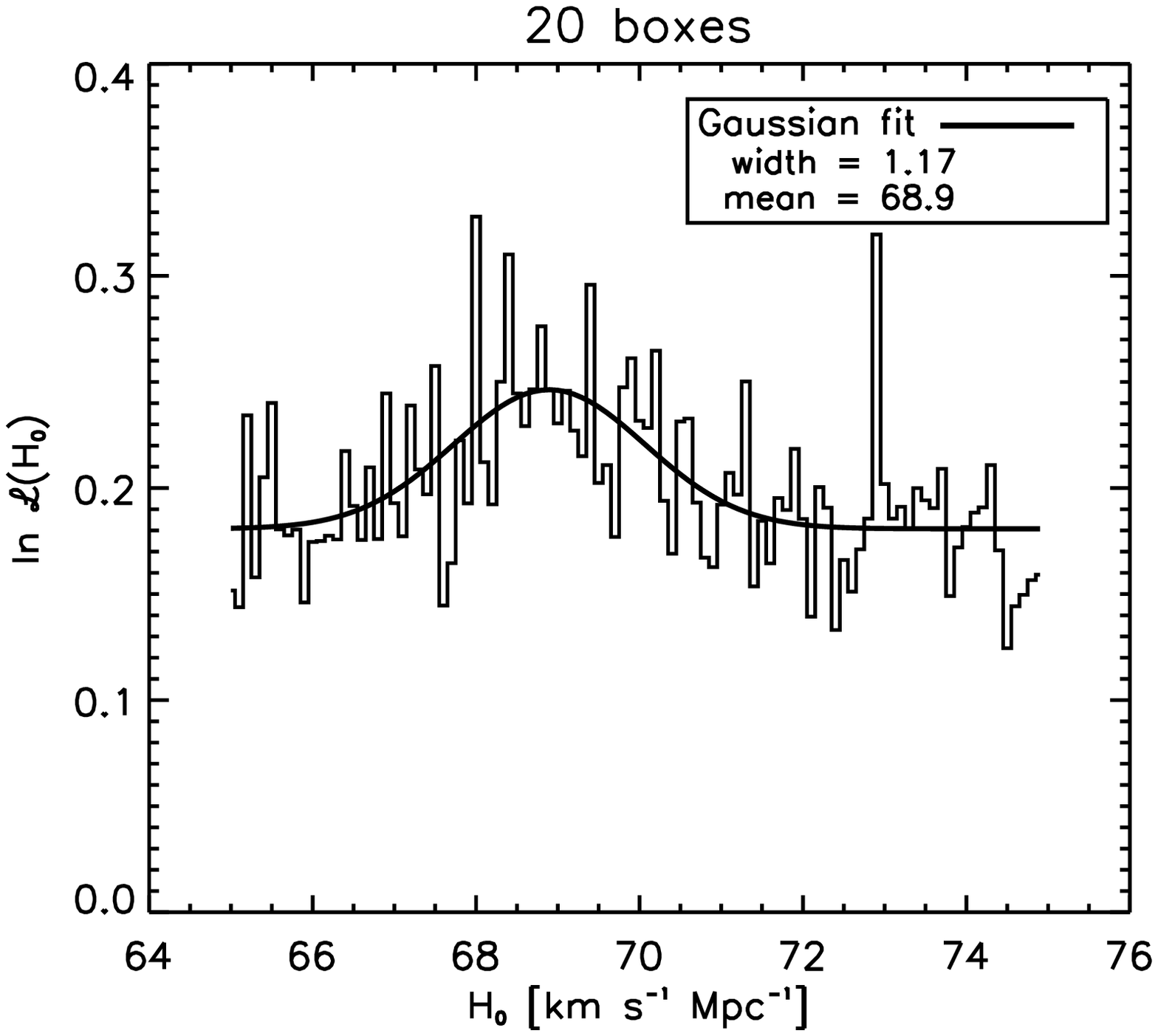}
     \includegraphics[width=2.5in]{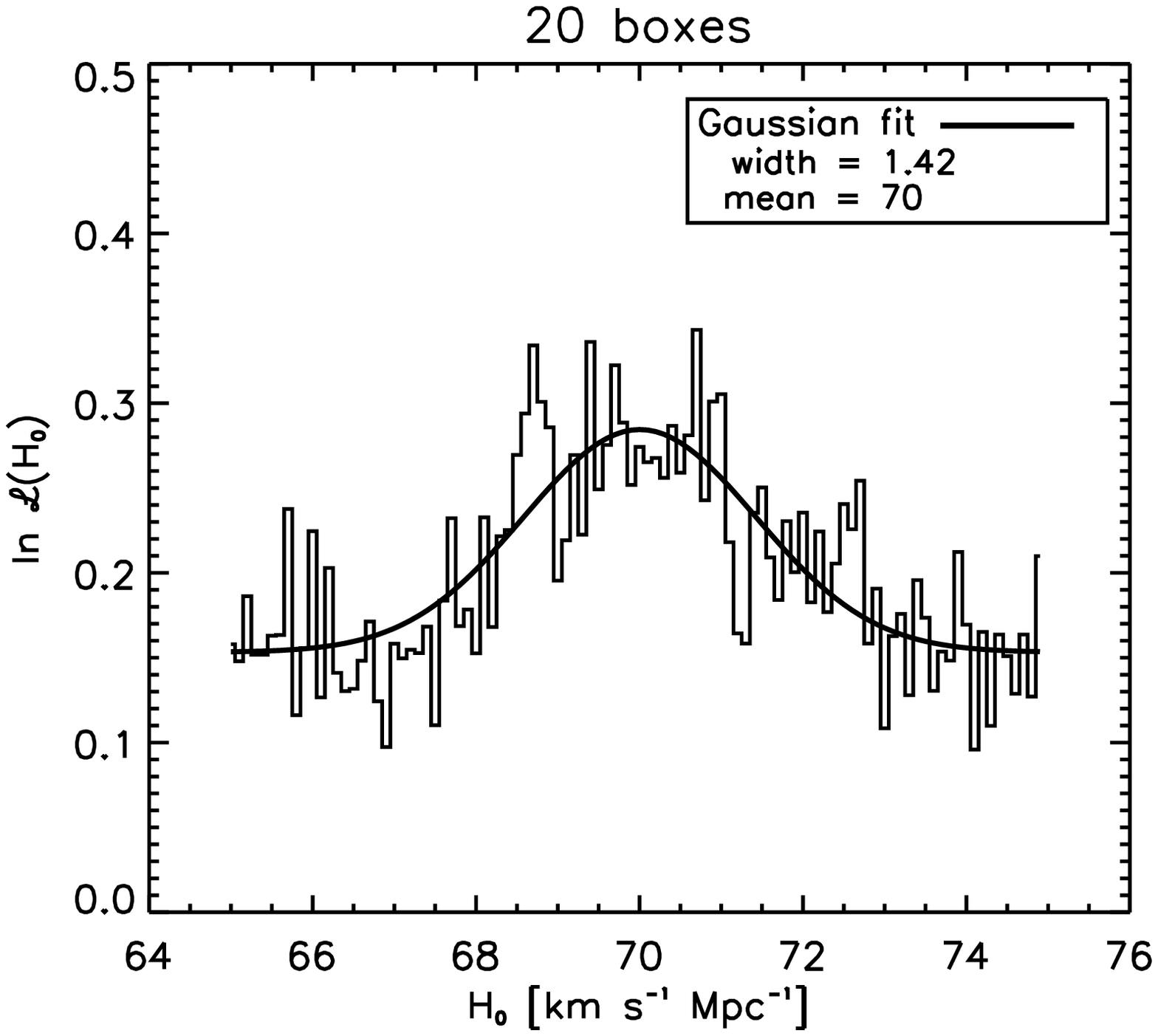}
   }
   \caption{Summed histograms for one synthetic interferometer,
   scaled to represent events at $z_{LISA} = 2z_{SDSS}$ (top row),
   $z_{LISA} = 3z_{SDSS}$ (second row), $z_{LISA} = 4z_{SDSS}$ (third
   row), and $z_{LISA} = 5z_{SDSS}$ (bottom row). Vertical axes are in arbitrary units.}
\label{fig:Aall}
\end{figure*}

The LISA distance error was taken into account using the same
method as before, except now scaled to $\Delta \ln(D_L) \approx 0.07
Az_{SDSS}$ for one synthetic interferometer. The results of these
realizations are shown in Fig.~\ref{fig:Aall}, where scalings of $A = $2, 3, 4,
and 5 were used.  

Results are summarized in Table~\ref{tab:A2}.  
It can be seen that these are much more realistic than the
lower-redshift realizations, with typical boxes containing hundreds of
EMRI host tracer galaxies.  Nevertheless, the final $H_0$ estimate remains
precise even for redshifts out to $z_{LISA} = 0.5$.  It should be
noted that since these error boxes are much larger and are still
limited to the northern galactic cap, significant overlapping of error
boxes occurs when scaling to high redshift, primarily for $A=5$. Therefore, the results for
these cases are not as reliable due to lack of independence.

\begin{figure*}[htbp]
   \centerline{
     \includegraphics[width=2.5in]{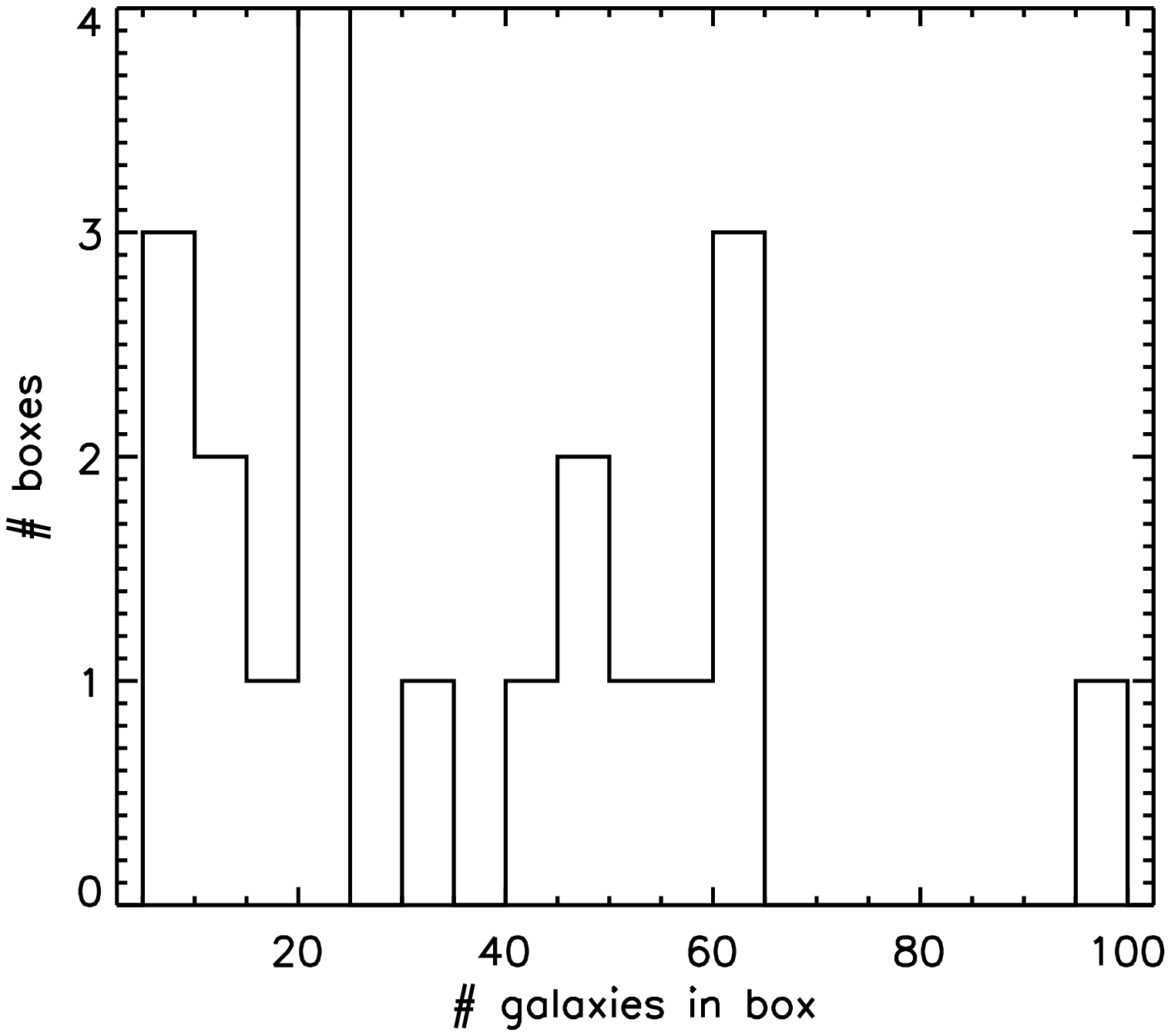}
     \includegraphics[width=2.5in]{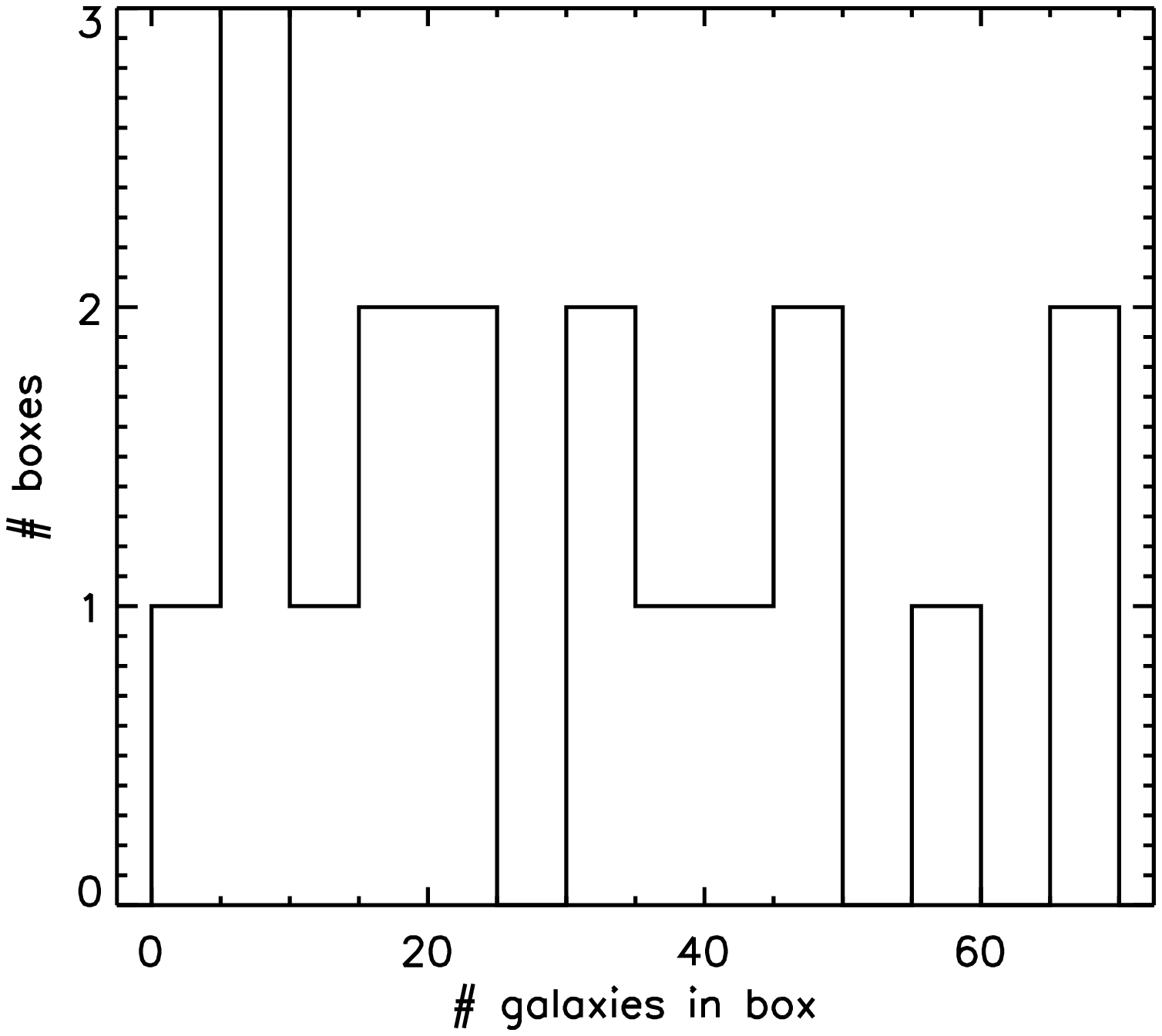}
     \includegraphics[width=2.5in]{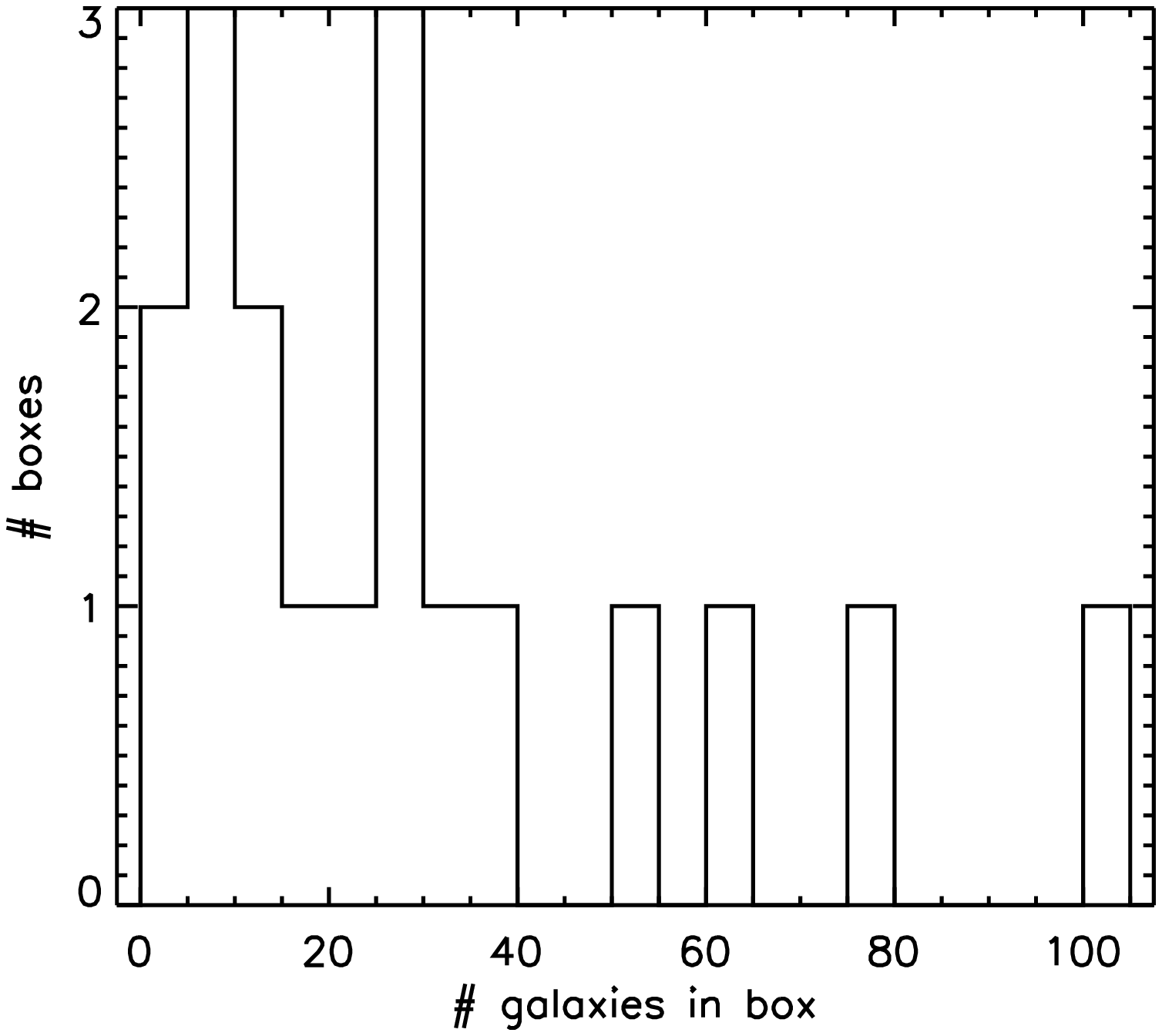}
   }\centerline{
     \includegraphics[width=2.5in]{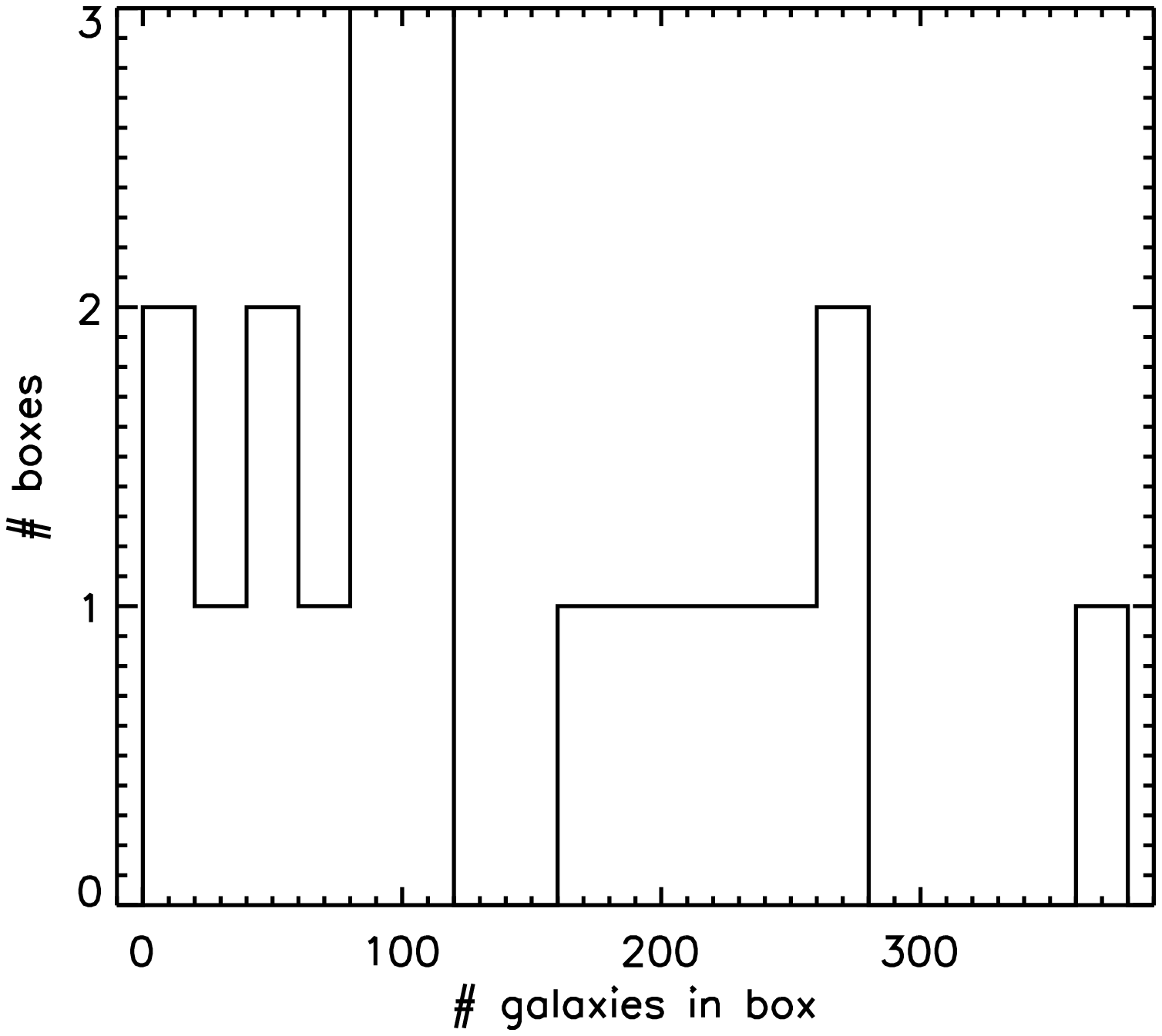}
     \includegraphics[width=2.5in]{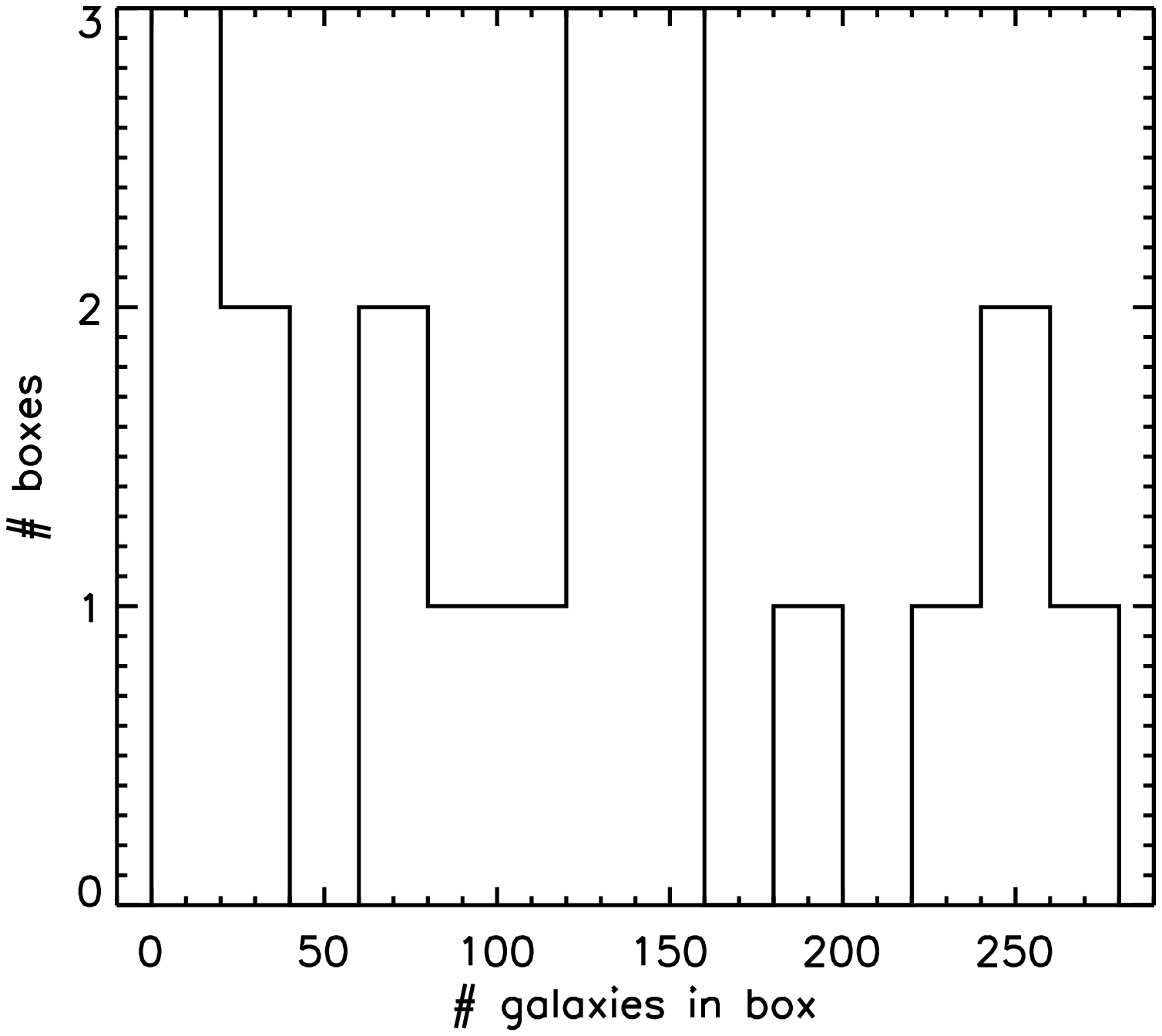}
     \includegraphics[width=2.5in]{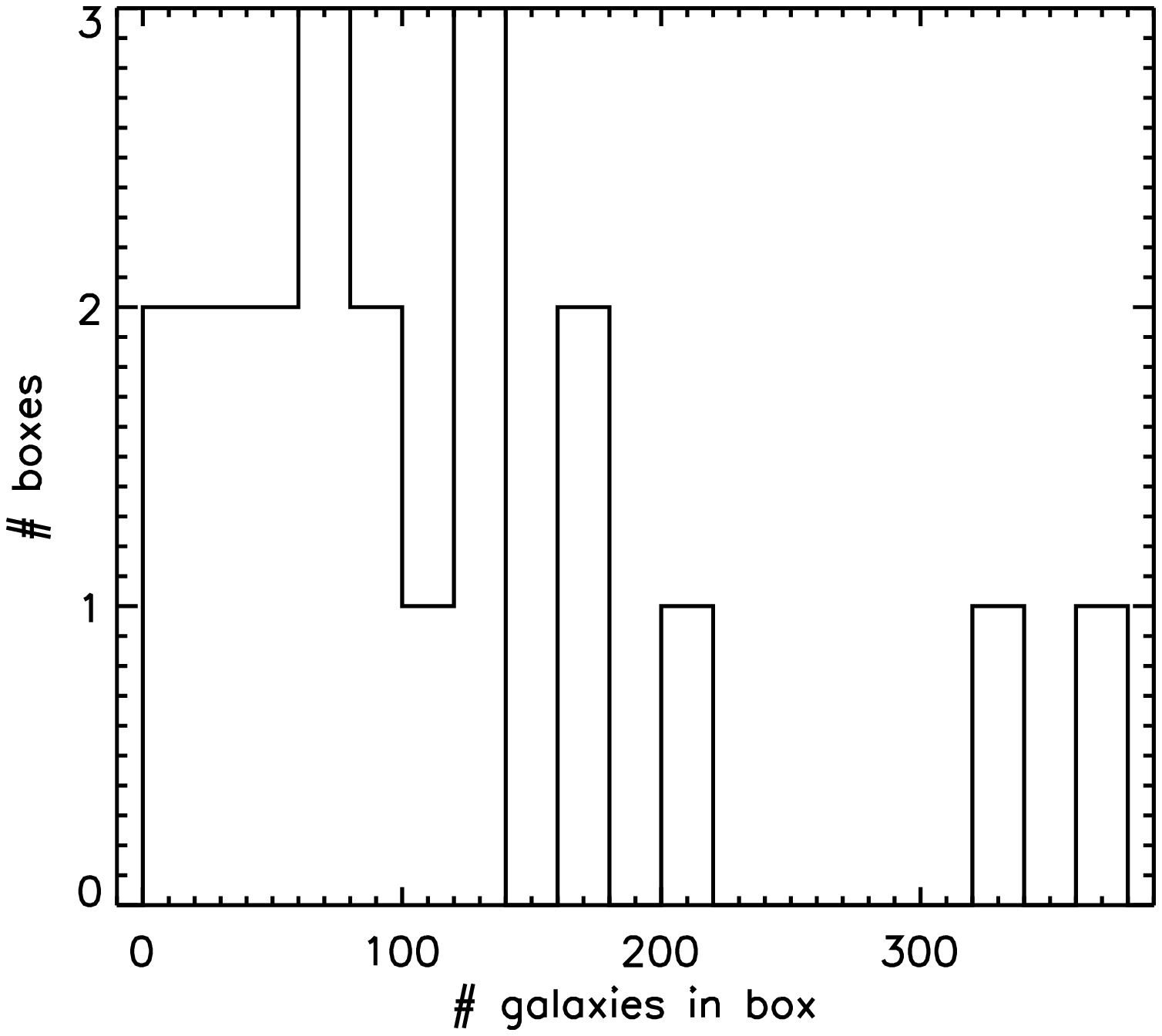}
   }\centerline{
     \includegraphics[width=2.5in]{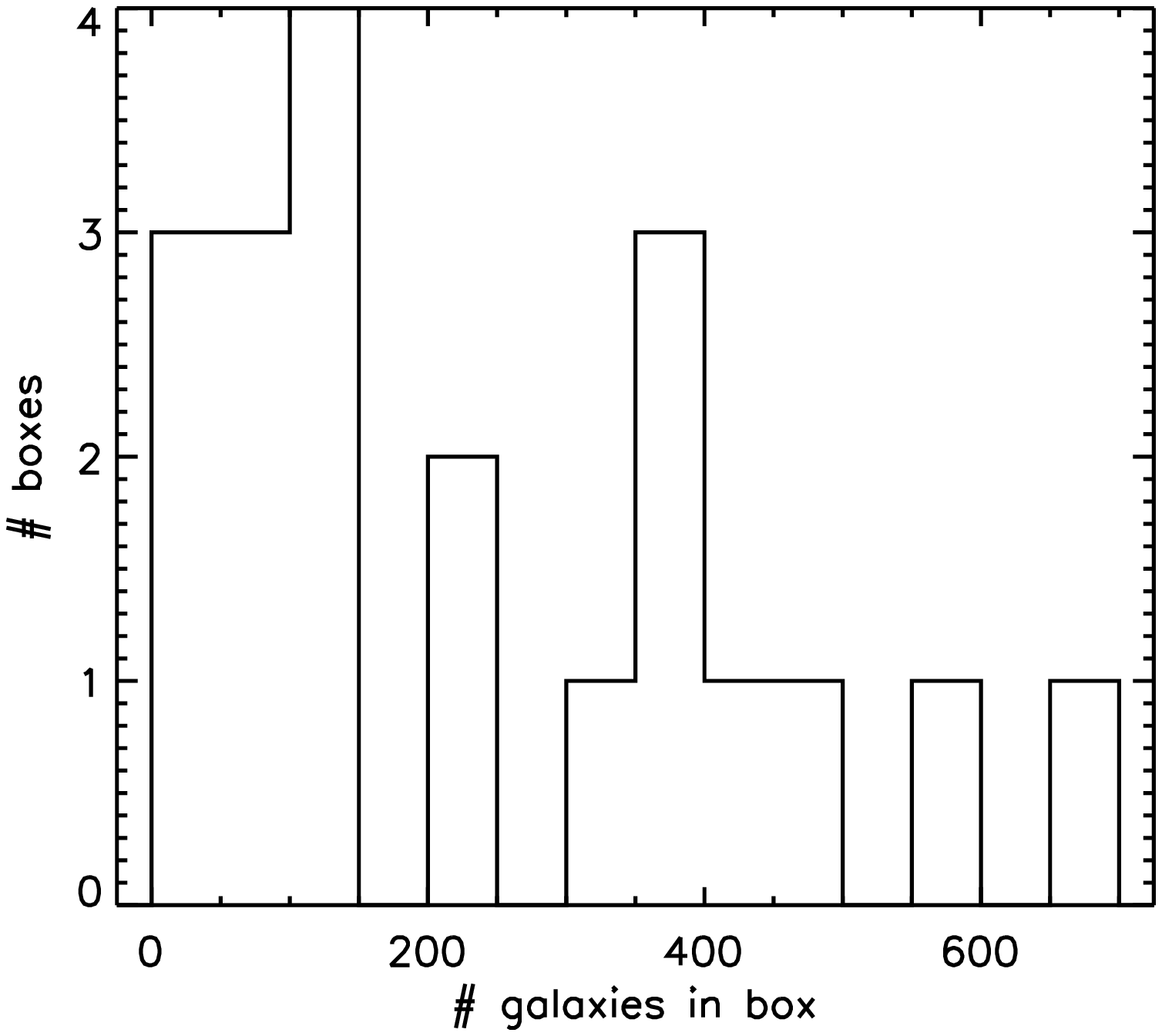}
     \includegraphics[width=2.5in]{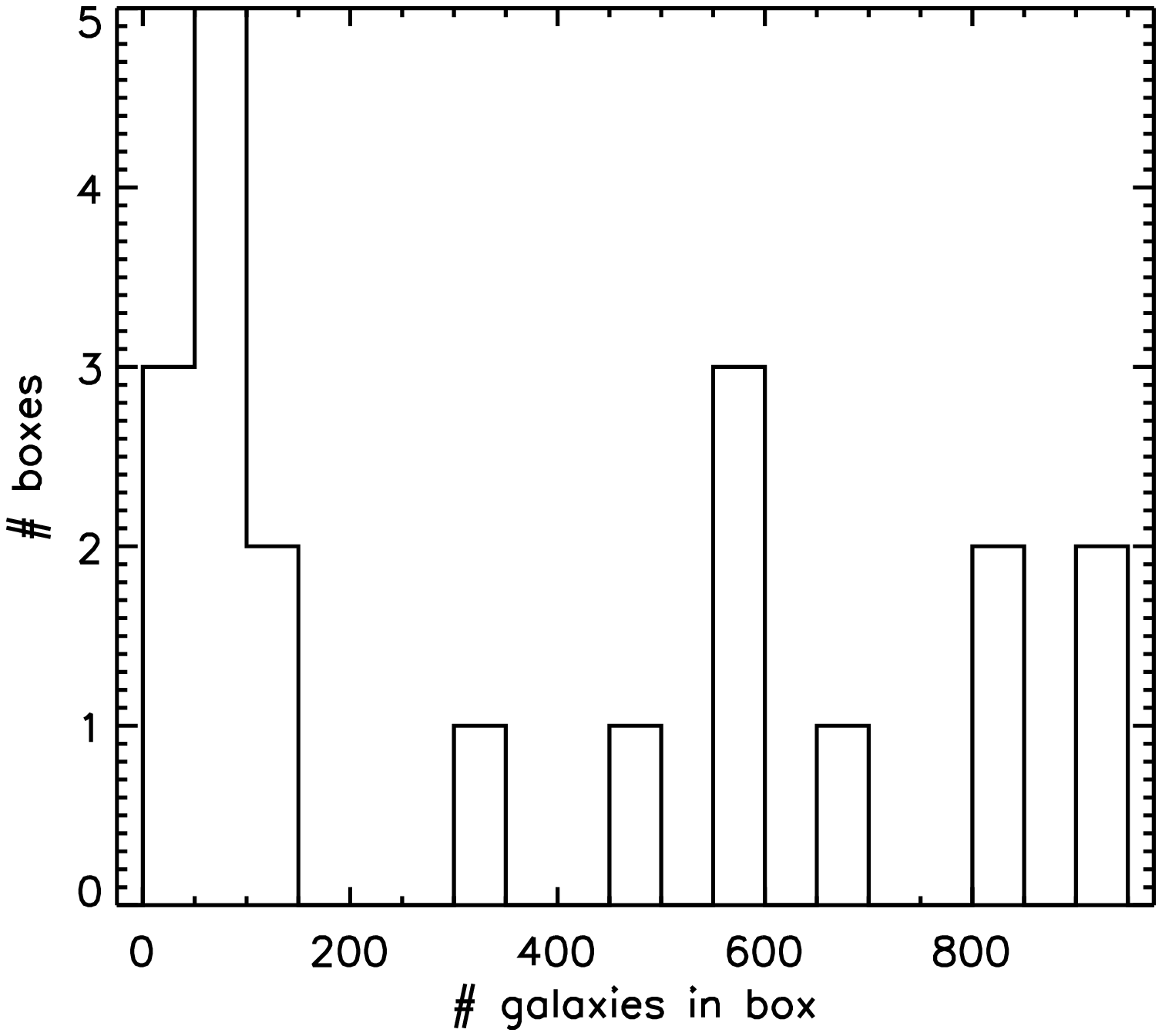}
     \includegraphics[width=2.5in]{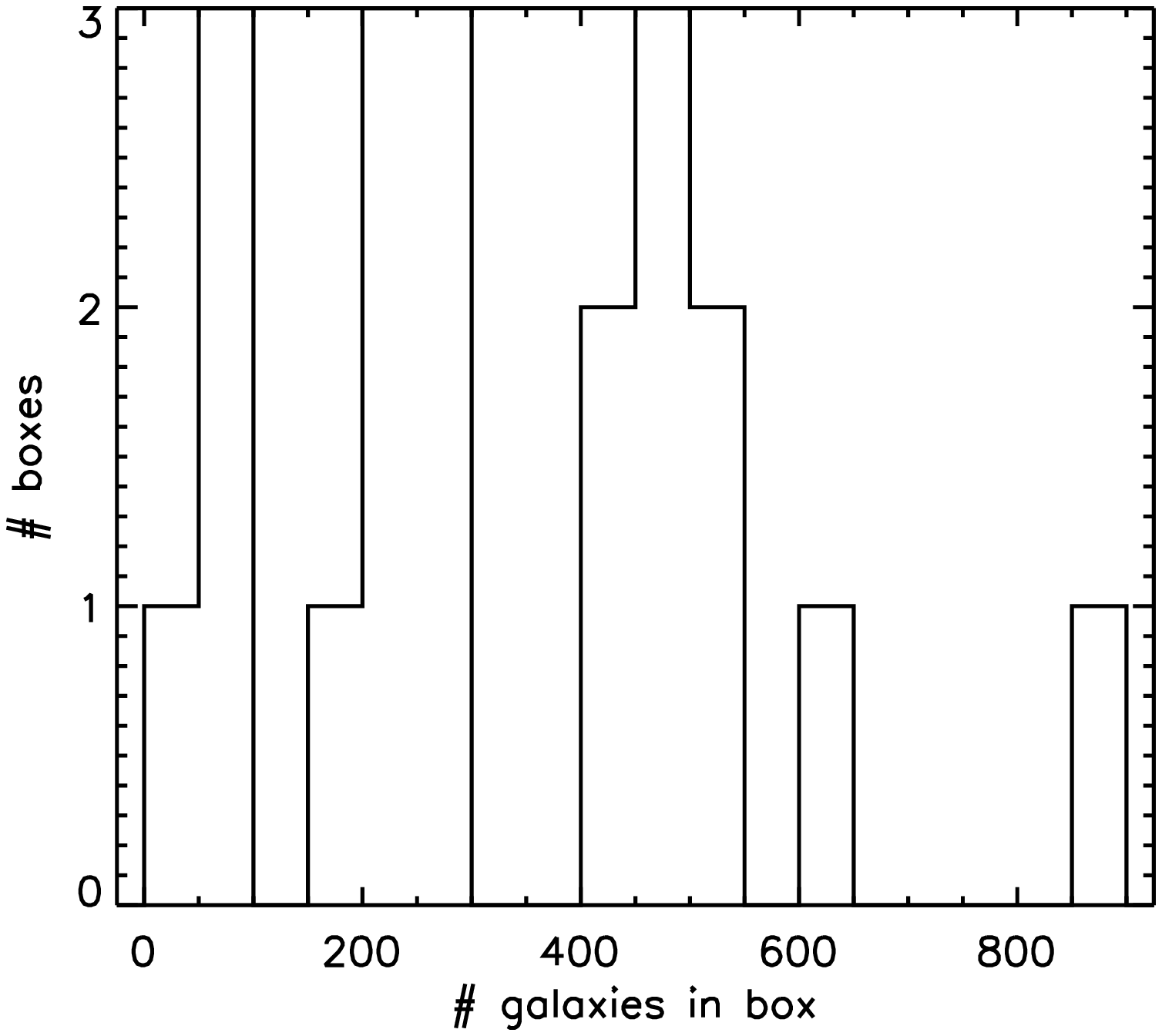}
   }\centerline{
     \includegraphics[width=2.5in]{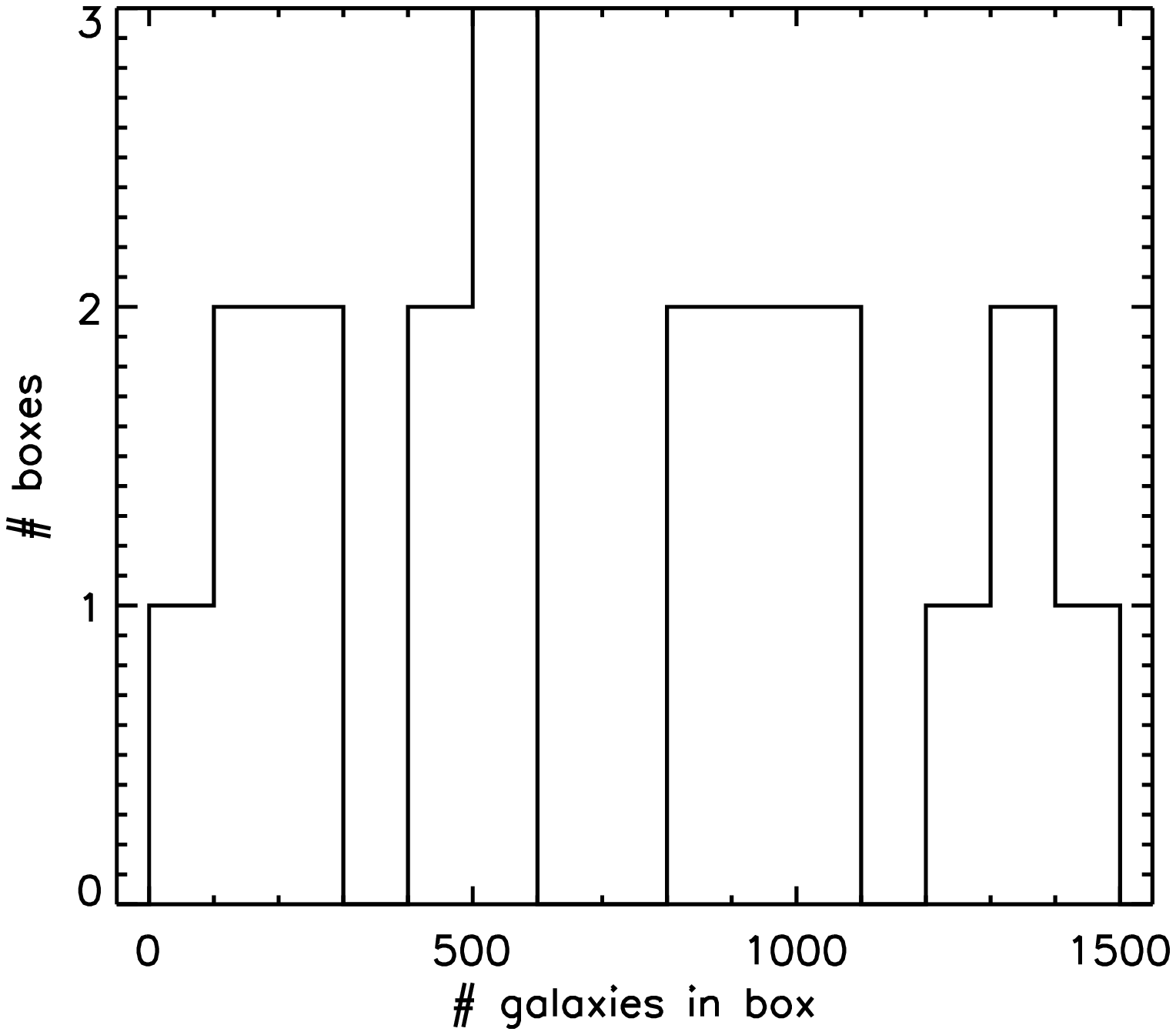}
     \includegraphics[width=2.5in]{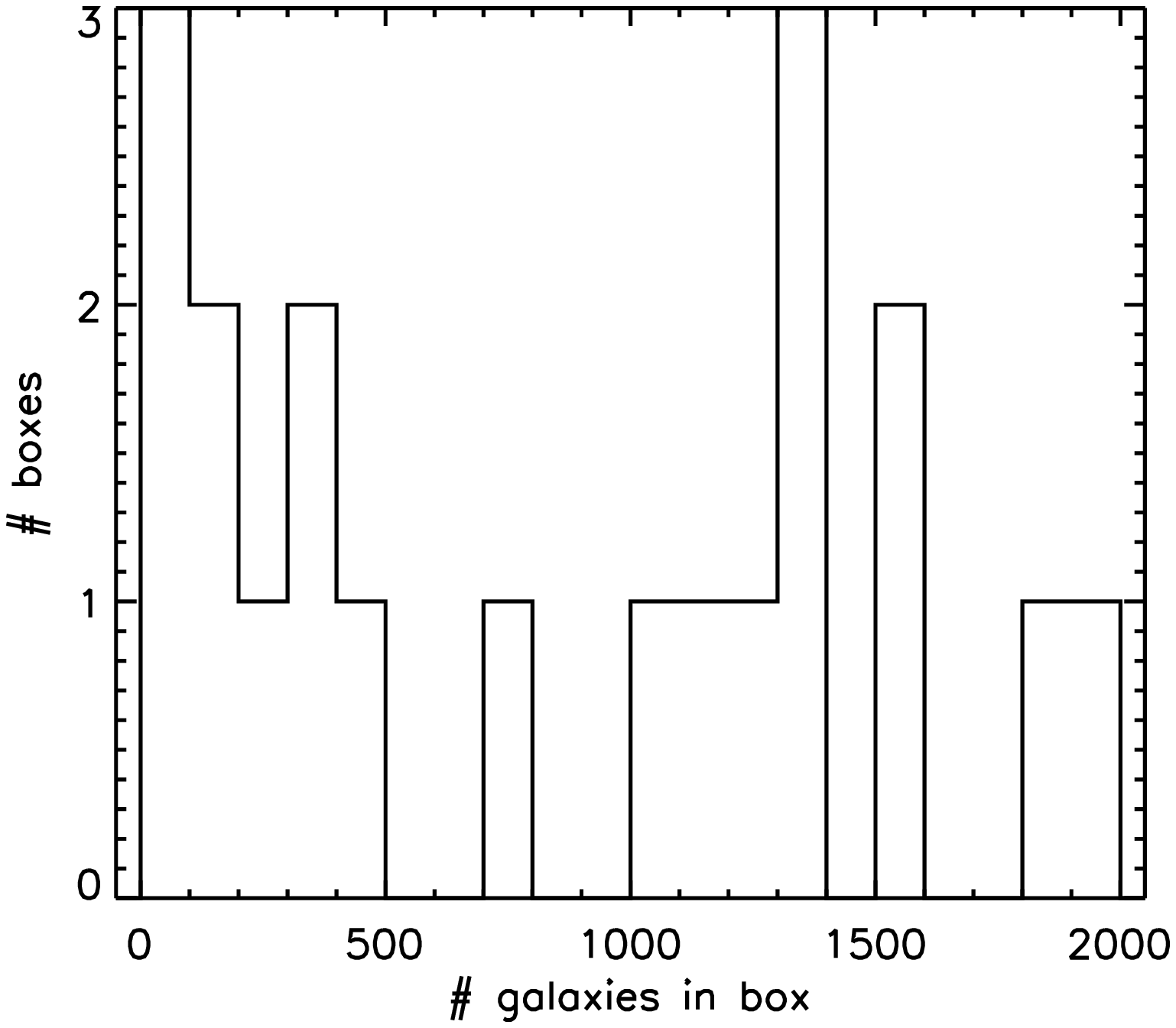}
     \includegraphics[width=2.5in]{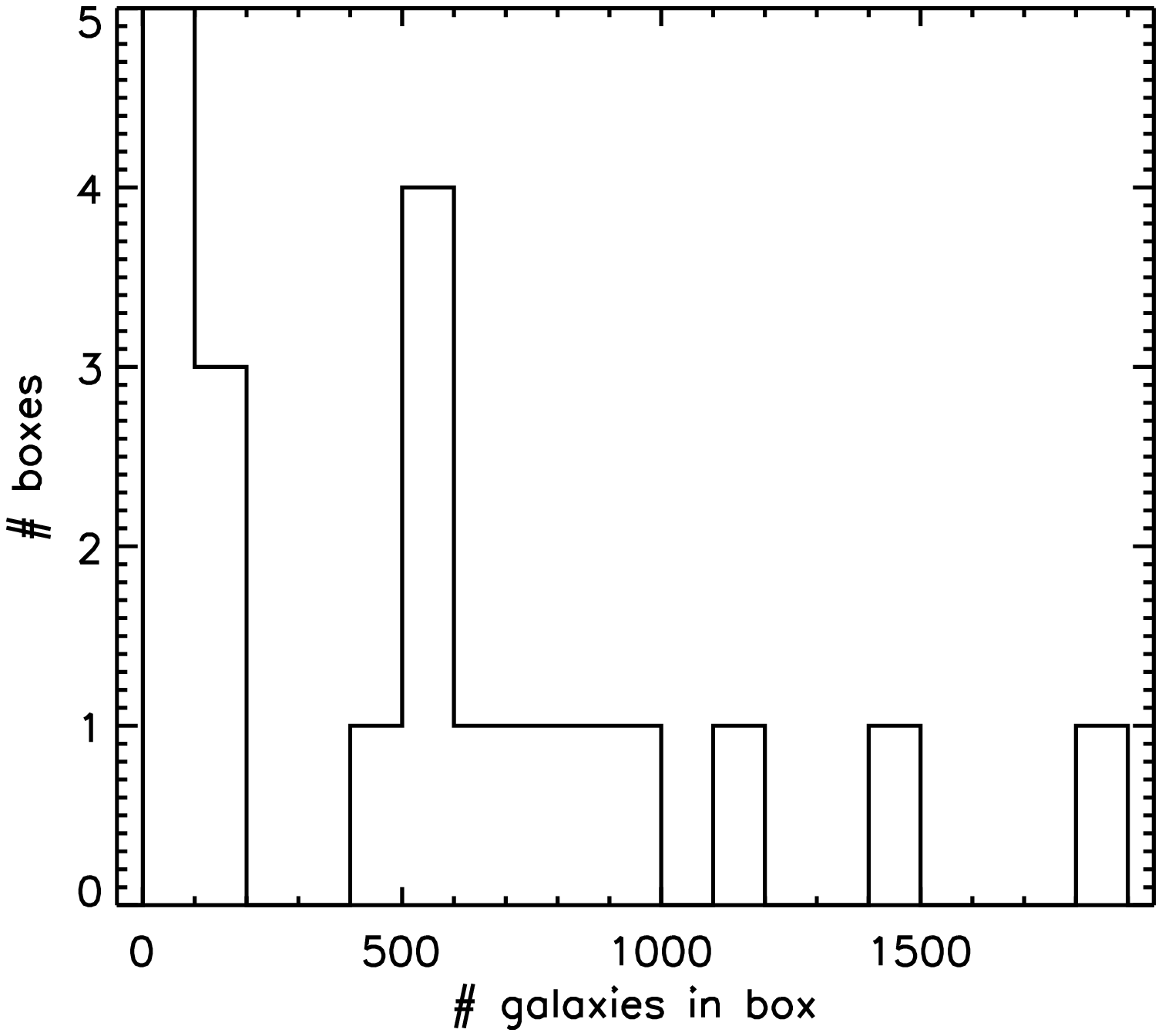}
   }
   \caption{Distributions of $N_j$ values for realizations in
   Figure~\ref{fig:Aall} (in the same order).}
\label{fig:Njs}
\end{figure*}

It should also be noted that even for these larger error boxes, there are still cases
where few galaxies ($<10$) are contained within a box, and
occasionally there will only be one to two galaxies.  These galaxies
contribute significantly to the final, summed
histograms. Figure~\ref{fig:Njs} shows histograms of the number of
galaxies contained in each box, that is, the distributions of $N_j$
values, for each of the plots in Fig.~\ref{fig:Aall}. 

The resulting errors in the mean ($\delta_{H_0} = |70 - mean|$) averaged over 25 realizations for
each value of $A$  are listed in Table~\ref{tab:errors}, all for the single synthetic
interferometer case. 

In conclusion, we have found that if LISA detects 20 or more EMRI
events to a redshift of $z \approx 0.5$,  galaxy surveys of the LISA error boxes are likely to yield a  reliable estimate of the Hubble constant to better than one percent precision.  A higher rate of EMRI events would permit estimates of  cosmic acceleration from the redshift-distance relation in this redshift range with considerably more precision than other known techniques.

\begin{table*}
\caption{Results for realizations in Fig.~\ref{fig:Aall}; starting from
  top left and going through each row left to right.}
\begin{tabular}{c c c c c c c}
\hline\hline
$A$ &Number of Boxes & Average $z_{SDSS}$ & Average $z_{LISA}$
& $\Sigma N_j$ & Width of Gaussian & Mean $H_0$\\
$$ &(Sources)& & 
& & & [km~s$^{-1}$~Mpc$^{-1}$]\\
\hline
2  & 20  & 0.115 & 0.231 & 712   & 0.97  & 69.74\\
2  & 18  & 0.103 & 0.206 & 532   & 0.89  & 70.03\\
2  & 18  & 0.105 & 0.211 & 525   & 0.54  & 69.81\\
3  & 20  & 0.111 & 0.333 & 2783  & 1.22  & 69.55\\
3  & 20  & 0.115 & 0.344 & 2454  & 1.11  & 69.97\\
3  & 20  & 0.088 & 0.263 & 2320  & 1.51  & 69.97\\
4  & 20  & 0.086 & 0.344 & 4794  & 0.93  & 69.75\\
4  & 20  & 0.108 & 0.434 & 7361  & 0.82  & 70.16\\
4  & 20  & 0.099 & 0.396 & 6668  & 1.21  & 70.69\\
5  & 20  & 0.112 & 0.558 & 14648 & 0.55  & 70.46\\
5  & 20  & 0.115 & 0.577 & 16929 & 1.17  & 68.90\\
5  & 20  & 0.088 & 0.442 & 11048 & 1.42  & 70.00\\
\hline\hline
\end{tabular}
\label{tab:A2}
\end{table*}

\begin{table*}
\caption{Errors in the mean $H_0$ averaged over 25 realizations, for
  one synthetic interferometer.}
\begin{tabular}{c  c}
\hline\hline
$A$ & $\delta_{H_0}$ [km~s$^{-1}$~Mpc$^{-1}$]\\
\hline
1  & 0.22\\
2  & 0.16\\
3  & 0.24\\
4  & 0.51\\
5  & 0.68\\
\hline\hline
\end{tabular}
\label{tab:errors}
\end{table*}

\section*{Acknowledgements}
Funding for the Sloan Digital Sky Survey (SDSS) and SDSS-II has been provided by the Alfred P. Sloan Foundation, the Participating Institutions, the National Science Foundation, the U.S. Department of Energy, the National Aeronautics and Space Administration, the Japanese Monbukagakusho, and the Max Planck Society, and the Higher Education Funding Council for England. The SDSS Web site is http://www.sdss.org/.

The SDSS is managed by the Astrophysical Research Consortium (ARC) for
the Participating Institutions. The Participating Institutions are the
American Museum of Natural History, Astrophysical Institute Potsdam,
University of Basel, University of Cambridge, Case Western Reserve
University, The University of Chicago, Drexel University, Fermilab,
the Institute for Advanced Study, the Japan Participation Group, The
Johns Hopkins University, the Joint Institute for Nuclear
Astrophysics, the Kavli Institute for Particle Astrophysics and
Cosmology, the Korean Scientist Group, the Chinese Academy of Sciences
(LAMOST), Los Alamos National Laboratory, the Max-Planck-Institute for
Astronomy (MPIA), the Max-Planck-Institute for Astrophysics (MPA), New
Mexico State University, Ohio State University, University of
Pittsburgh, University of Portsmouth, Princeton University, the United
States Naval Observatory, and the University of Washington.


\begin{thebibliography}{5}
\bibitem[Schutz(1986)]{1986Natur.323..310S} B.~F. Schutz, \nat {\bf 323}, 
310 (1986).\\
\bibitem[Dalal et al.(2006)]{2006PhRvD..74f3006D} N. Dalal, D.~E. Holz, 
S.~A. Hughes, and B. Jain, \prd {\bf 74}, 063006 (2006).\\
\bibitem[Holz \& Hughes(2005)]{2005ApJ...629...15H} D.~E. Holz and
  S.~A. Hughes, \apj {\bf 629}, 15 (2005).\\
\bibitem[Flanagan \& Hughes(2005)]{flan}
E. Flanagan and S. Hughes, New J. Phys. {\bf 7}, 204 (2005).\\
\bibitem[Hogan(2007)]{2007arXiv0709.0608H} C.~J. Hogan, in Frontiers
  of Astrophysics: A Celebration of NRAO's 50th Anniversary,
  Charlottesville, 2007, edited by A.~H. Bridle, J.~J. Condon and
  G.~C. Hunt
  (to be published); Report No. arXiv:0709.0608v2, 2007.\\
\bibitem[Amaro-Seoane et al.(2007)]{2007CQGra..24..113A}
  P. Amaro-Seoane, J.~R. Gair, M. Freitag, M.~C. Miller, I. Mandel,
  C.~J. Cutler, and S. Babak, Class. and Quantum Grav. {\bf 24},
  113 (2007).\\
\bibitem[LISA(2006)]{LISA} See {\it Laser Interferometer Space Antenna: 6th
  International LISA Symposium}, edited by S.~M. Merkowitz and J.~C. Livas (Greenbelt, 2006) [AIP Conf. Proc. {\bf 873}].\\ 
\bibitem[Hu(2005)]{Hu} W. Hu, in {\it ASP Conf. Ser. {\bf 399}}, edited by S.~C. Wolff
and T.~R. Lauer (San Francisco, 2005), p. 215.\\
\bibitem[Knox(2006)]{Knox} L. Knox, \prd {\bf 73}, 023503 (2006).\\
\bibitem[Hughes \& Holz(2003)]{2003CQGra..20S..65H} S.~A. Hughes and
  D.~E. Holz, Class. and Quantum Grav. {\bf 20}, 65 (2003).\\
\bibitem[Trias \& Sintes(2007)]{2007arXiv0707.4434T} M. Trias and
  A.~M. Sintes, Phys. Rev. D. (submitted); Report No. arXiv:0707.4434, 2007.\\
\bibitem[Berti et al.(2005)]{2005PhRvD..71h4025B} E. Berti, A. Buonanno, 
and C.~M. Will, \prd {\bf 71}, 084025 (2005).\\
\bibitem[Vecchio(2004)]{2004PhRvD..70d2001V} A. Vecchio, \prd {\bf 70}, 
042001 (2004).\\
\bibitem[Hughes(2002)]{2002MNRAS.331..805H} S.~A. Hughes, Mon. Not. R. Astron. Soc. 
{\bf 331}, 805 (2002).\\
\bibitem[Cutler(1998)]{1998PhRvD..57.7089C} C. Cutler, \prd {\bf 57}, 
7089 (1998).\\
\bibitem[Sesana et al.(2004)]{2004ApJ...611..623S} A. Sesana, F. Haardt, 
P. Madau, and M. Volonteri, \apj {\bf 611}, 623 (2004).\\
\bibitem[Volonteri(2006)]{2006AIPC..873...61V} M. Volonteri, AIP
  Conf. Proc. {\bf 873}, 61 (2006).\\
\bibitem[de Freitas Pacheco et al.(2006)]{2006PhRvD..74b3001D} J.~A. de Freitas 
Pacheco, C. Filloux, and T. Regimbau, \prd {\bf 74}, 023001 (2006).\\
\bibitem[Gair et al.(2004)]{2004CQGra..21S1595G} J.~R. Gair, L. Barack, 
T. Creighton, C. Cutler, S.~L. Larson, E.~S. Phinney, and
M. Vallisneri, Class. and Quantum Grav. {\bf 21}, 1595 (2004).\\
\bibitem[Barack \& Cutler(2004)]{2004PhRvD..69h2005B} L. Barack and 
C. Cutler, \prd {\bf 69}, 082005 (2004).\\
\bibitem[Hopman(2006)]{2006AIPC..873..241H} C. Hopman, AIP
  Conf. Proc. {\bf 873}, 241 (2006).\\
\bibitem[Hopman \& Alexander(2006)]{2006ApJ...645L.133H}  C. Hopman
  and T. Alexander, Astrophys. J. Lett. {\bf 645}, L133 (2006).\\
\bibitem[Dotti et al.(2006)]{2006MNRAS.372..869D} M. Dotti,
  R. Salvaterra, A. Sesana, M. Colpi, and F. Haardt,
  Mon. Not. R. Astron. Soc. {\bf 372},
  869 (2006).\\
\bibitem[Milosavljevi{\'c} \& Phinney(2005)]{2005ApJ...622L..93M} 
M. Milosavljevi{\'c} and E.~S. Phinney, Astrophys. J. Lett. {\bf 622}, L93 (2005).\\
\bibitem[E. S. Phinney, in prep.]{phinney} E.~S. Phinney (in
  preparation).\\
\bibitem[Armitage \& Natarajan(2002)]{2002ApJ...567L...9A}  P.~J. Armitage 
and P. Natarajan, Astrophys. J. Lett. {\bf 567}, L9 (2002).\\
\bibitem[Kocsis et al.(2006)]{2006ApJ...637...27K} B. Kocsis, Z. Frei, 
Z. Haiman, and K. Menou, \apj {\bf 637}, 27 (2006).\\
\bibitem[Kocsis et al.(2007)]{2007PhRvD..76b2003K} B. Kocsis, Z. Haiman, 
K. Menou, and Z. Frei, \prd {\bf 76}, 022003 (2007).\\
\bibitem[Kocsis et al.(2007)]{2007arXiv0712.1144K} B. Kocsis, Z. Haiman, and K. Menou, \apj (submitted); Report No. arXiv:0712.1144, 2007.\\
\bibitem[Adelman-McCarthy et al. (2008)]{DR6} J. Adelman-McCarthy {\it et
  al.}, Astrophys. J. Suppl. Ser. (submitted).\\ 
\bibitem[Stoughton et al.(2002)]{2002AJ....123..485S} C. Stoughton {\it et
  al.}, Astron. J. {\bf 123}, 485 (2002).\\ 
\bibitem[Strauss et al.(2002)]{2002AJ....124.1810S} M.~A. Strauss {\it et
  al.}, Astron. J. {\bf 124}, 1810 (2002).\\ 
\bibitem[Gunn et al.(2006)]{2006AJ....131.2332G} J.~E. Gunn {\it et al.}, 
Astron. J. {\bf 131}, 2332 (2006).\\ 
\bibitem[Gunn et al.(1998)]{camera} J.~E. Gunn {\it et al.}, Astron. J.
{\bf 116}, 3040 (1998).\\ 
\bibitem[York et al.(2000)]{2000AJ....120.1579Y} D.~G. York {\it et al.}, 
Astron. J. {\bf 120}, 1579 (2000).\\ 
\bibitem[Fukugita et al. (1996)]{filters} M. Fukugita, T. Ichikawa,
   J.~E. Gunn, M. Doi, K. Shimasaku, and D.~P. Schneider,
   Astron. J. {\bf 111}, 1748 (1996).\\ 

\end{thebibliography}
\end{document}